\documentclass[oldversion]{aa}     
\def \th {\thinspace}
\def \arcmin {\hbox{$^\prime$}}

\def\approxgt{\mathrel{\hbox{\rlap{\lower.55ex \hbox {$\sim$}} \kern-.3em \raise.4ex \hbox{$>$}}}}
\def\lesssim{\mathrel{\hbox{\rlap{\lower.55ex \hbox {$\sim$}} \kern-.3em \raise.4ex \hbox{$<$}}}}
\def\approxlt{\mathrel{\hbox{\rlap{\lower.55ex \hbox {$\sim$}} \kern-.3em \raise.4ex \hbox{$<$}}}}
\def \degmark {^\circ}
\def \sun {\hbox {$\odot$}}
\usepackage{graphicx}
\begin{document}

\title{Spectral investigations of the nature of the Sco\th X-1 like sources}

\author{M. J. Church\inst{1}
\and A. Gibiec\inst{2}
\and M. Ba\l uci\'nska-Church\inst{1}
\and N. K. Jackson\inst{1}}
\institute{School of Physics and Astronomy, University of Birmingham,
           Birmingham, B15 2TT, UK\\
\and
          Astronomical Observatory, Jagiellonian University,
          ul. Orla 171, 30-244 Cracow, Poland.\\}
\offprints{mjc@star.sr.bham.ac.uk}
\date{Received February 8 2012; Accepted August 15 2012}
\titlerunning{A model for Sco\th X-1 like sources}
\authorrunning{Church et al.}

\abstract{
We present results of spectral investigations of the Sco\th X-1 like Z-track sources Sco\th X-1, GX\th 349+2
and GX\th 17+2 based on {\it Rossi-XTE} observations determining spectral evolution along the Z-track.
Results are obtained for a spectral model describing an extended accretion disk corona.
The results are compared with previous results for the Cyg\th X-2 like group: Cyg\th X-2, GX\th 340+0 and
GX\th 5-1 and a general model for the Z-track sources proposed. On the normal branch, the Sco-like and Cyg-like
sources are similar, the results indicating an increase of mass accretion rate $\dot M$ between soft and hard
apex, not as in the standard view that this increases monotonically around the Z. In the Cyg-like sources,
increasing $\dot M$ causes the neutron star temperature $kT$ to increase from $\sim$1 to $\sim$2 keV. At the lower
temperature, the radiation pressure is small, but at the higher $kT$, the emitted flux of the neutron star is
several times super-Eddington and the high radiation pressure disrupts the inner disk and launches the relativistic
jets observed on the upper normal and horizontal branches. In the Sco-like sources, the main physical difference
is the high $kT$ of more than 2 keV
on all parts of the Z-track suggesting jets are always possible, even on the flaring branch.
The flaring branch in the Cyg-like sources is associated with a release of energy on the neutron star consistent
with unstable nuclear burning.
The Sco-like sources are very different as flaring appears to be a {\it combination} of 
unstable burning and an increase of $\dot M$ which makes flaring much stronger.
Analysis of 15 years of {\it RXTE} ASM data on all 6 classic Z-track sources
shows the high rate and strength of flaring in the Sco-like sources suggesting that the continual release
of energy heats the neutron star causing the high $kT$. Analysis of a Sco\th X-1 observation with unusually
little flaring supports this. GX\th 17+2 appears to be transitional between the Cyg and Sco-like types. 
Our results do not support the suggestion that Cyg or Sco-like nature is determined
by the luminosity.
}
\maketitle

\section{Introduction}

The Z-track sources form the brightest group of Low Mass X-ray Binaries (LMXB)
containing a neutron star, with luminosities at or above the Eddington limit. The sources
display three tracks in hardness-intensity diagrams: the Horizontal Branch (HB), Normal Branch (NB) 
and Flaring Branch (FB) (Hasinger \& van der Klis 1989) showing that major physical changes
take place within the sources. However, the nature of these changes has not been understood
and is essential to an understanding of LMXB in general. It has been widely thought that the 
physical changes are caused by change of the mass accretion rate along the Z-track 
(Priedhorsky et al. 1986), assumed to increase in the direction HB$\rightarrow$NB$\rightarrow$FB, 
however, the evidence for this is rather limited (below). Moreover, the variation 
of X-ray intensity does not suggest this, increasing on the HB, but then decreasing on the NB.

It was apparent that the sources fell into two groups (Kuulkers et al. 1997; Hasinger 
\& van der Klis 1989): firstly, the 
Cyg-like sources Cygnus\th X-2, GX\th 5-1 and GX\th 340+0 displaying full Z-tracks with HB, NB and FB 
but having weak flaring. Secondly in the Sco\th X-1 like group: Sco\th X-1, GX\th 349+2 and GX\th 17+2 
there is hardly any HB but flaring is strong and frequent with large increases of X-ray intensity. 
These substantial differences have not been understood.

The sources are important because of the observation of relativistic jets in the radio, notably 
the charting of outward motion of material from Sco\th X-1 by Fomalont et al. (2001) at a 
velocity $v/c$ $\sim$0.45. As jets are seen predominantly in the HB (e.g. Penninx 1989),
X-ray observations allow us to compare conditions at the inner disk when 
jets are formed and are not formed, allowing us to investigate jet launching observationally as 
opposed to theoretically, aiming to understand the disk-jet connection.

Since the discovery of the sources as a class with three basic states (Schulz et al. 1989)
there has been relatively little progress in understanding them. Extensive timing studies have 
been carried out (van der Klis et al. 1987; Hasinger et al. 1989) revealing complex behaviour 
of quasi-periodic oscillations (QPO) at low frequencies ($\sim$6 Hz) and at kilohertz frequencies;
see review of van der Klis (2003).
Spectral investigations may be more likely to provide an explanation, as revealing
physical changes in the emission components during motion along the Z-track, but have been 
hindered by lack of agreement on spectral models. 

\subsection{The extended ADC}

The broadband spectra exhibit both non-thermal and thermal emission, which have
been interpreted using two very different physical models as have the spectra of LMXB in general. 
In the Eastern model (Mitsuda et al. 1989), non-thermal emission originates in a small central 
region and thermal emission from the inner disk. It has been widely applied to the Z-track sources,
however, there is no generally accepted physical interpretation of the resulting parameter
variations.

The Extended Accretion Disk Corona model (Church \& Ba\l uci\'nska-Church 1995, 2004) was 
developed out of the original Western model in which blackbody emission is from the neutron star and 
non-thermal emission is Comptonization from an ADC above the disk. The extended nature of the 
accretion disk corona (ADC) 
was revealed by work on the dipping class of LMXB, sources having high inclination such that decreases 
of X-ray intensity are seen in every binary orbit, due to absorption in the bulge in the
outer disk where the accretion flow impacts (White \& Swank 1982; Church et al. 1997). 

The dipping sources strongly constrain emission models since spectra 
of non-dip and several levels of dipping must all be fitted. The gradual removal in dips
of the dominant Comptonized emission over a time of order 100 seconds argues strongly for the
emission being extended. Measurements of the dip ingress time in the dipping sources show that the ADC has 
radial sizes $R$ of 10\th 000 to 700\th 000 km depending on luminosity, and is thin, i.e with 
a height $H$ having $H/R$ $<<$ 1: a thin corona above a thin disk (Church \& Ba\l uci\'nska-Church 2004). 
The spectral evolution in dipping shows that the Comptonized emission is removed gradually by an extended absorber 
having increasing overlap with the extended source, again demonstrating the extended nature of the ADC.
The Extended ADC model gave very good fits to the complex spectral evolution in 
dipping in many observations of the dipping LMXB (Church et al. 1997; 1998a,b, 2005; Ba\l uci\'nska-Church
et al. 1999, 2000; Smale et al. 2001; Barnard et al. 2001). In addition it was able to fit the spectra
of all types of LMXB in a survey made with {\it ASCA} (Church \& Ba\l uci\'nska-Church 2001) and so
capable of describing sources of all inclinations. 

Recently support for the extended ADC came from the {\it Chandra} grating results of
Schulz et al. (2009) by an independent technique. Precise grating measurements revealed a wealth
of emission lines of highly excited states (many H-like ions) such as Ne X, Mg XII, Si XIV,
S XIV, S XVI, Fe XXV and Fe XXVI. The line widths seen as Doppler shifts due to orbital motion 
indicated emission in a hot ADC at radial positions between 18\th 000 and 110\th 000 km 
in good agreement with the overall ADC size from dip ingress timing. 

In multi-wavelength observations of Cygnus\th X-2 in 2009 (Ba\l uci\'nska-Church et al. 2011)
a series of absorption edges below 1.4 keV remarkably had no detectable increase 
in depth during dips, showing that part of the continuum emission had to be extended
so that part was not covered, while covered emission below 1.4 keV was completely removed.

Thus the evidence favours an extended ADC and this model has provided 
an explanation of the Cygnus\th X-2 sub-group (below)
which seems convincing
and has the potential to lead to a better understanding of the sources and jet formation. 
In the present work we test the hypothesis that the model can provide a physically
reasonable explanation of the Sco\th X-1 like sub-group.

\subsection{A model for the Cygnus\th X-2 like Z-track sources}

We previously applied the Extended ADC model to high-quality {\it Rossi-XTE} data on the Cyg-like 
sources GX\th 340+0 (Church et al. 2006; Paper I), GX\th 5-1 (Jackson et al. 2009; Paper II) and Cygnus\th X-2
(Ba\l uci\'nska-Church et al. 2010; Paper III). In all cases the model fitted the spectra well at all positions
on the Z-track, and clearly suggested a physical explanation for the Z-track phenomenon. All three
sources behaved in the same way showing the explanation to be consistent. In this model, the Soft Apex 
between the NB and FB has a low mass accretion rate $\dot M$ and so a low X-ray intensity. Moving
along the NB towards the Hard Apex, there is a strong increase in luminosity, 90\% of which is due to
the ADC Comptonized emission and it is difficult to see how this can happen without an increase of mass accretion rate.
The resultant increase in $kT_{\rm BB}$ from $\sim$1 keV at the soft apex to 
$\sim$2 keV at the hard apex and higher values still on the HB results in a large increase in $T_{\rm BB}^4$ 
and so a large increase in the radiation pressure of the neutron star. 
The ratio $f/f_{\rm Edd}$: the flux emitted by unit area of the star $f$ as a fraction of the Eddington
value ($L_{\rm Edd}/4\,\pi\,R_{\rm NS}^2$, where $R_{\rm NS}$ is the radius of the neutron star) increases 
from $\sim$0.3 at the soft apex 
to 3 times super-Eddington on the HB. The good correlation of high radiation pressure with the radio emission
of jets on this part of the Z-track led us to suggest that high radiation pressure disrupts the inner disk
deflecting the accretion flow vertically so launching the jets.

On the HB $L_{\rm ADC}$ decreases suggesting $\dot M$ decreases while $kT_{\rm BB}$ continues to rise.
It is not definite what happens: $\dot M$ may be decreasing in the disk affecting $L_{\rm ADC}$ 
but not yet reaching the neutron star;
alternatively, the high radiation pressure may be disrupting the disk out to larger radial positions
so reducing the Comptonized emission.

In flaring, $L_{\rm ADC}$ was constant, suggesting that $\dot M$ was constant, but
there is a strong increase in $L_{\rm BB}$. If $\dot M$ does not change, 
flaring must be due to an additional energy source on the neutron star, 
i.e. unstable nuclear burning. Support for this came from measured values of the mass accretion rate per 
unit emitting area on the neutron star $\dot m$, the critical parameter in the theory of stable/unstable 
burning on the surface of the neutron star (e.g. Bildsten 1998). The theory describes various r\'egimes of
stable and unstable nuclear burning over the wide range of luminosities
found in LMXB, including unstable burning at lower luminosities seen as X-ray bursts
but is also applicable at the luminosities of the Z-sources. In all three sources, 
$\dot m$ at the onset of flaring (the soft apex) agreed with the theoretical critical value below which
there is unstable helium burning.

\subsection{A monotonic variation of mass accretion rate ?}

This model for the Cygnus\th X-2 like sources with $\dot M$ increasing on the NB and constant 
on the FB does not support the standard view that $\dot M$ increases monotonically HB$\rightarrow$NB$\rightarrow$FB.
In a multi-wavelength campaign on Cyg\th X-2 (Hasinger et al. 1990),  an increase of UV emission (Vrtilek et al. 1990) 
in what appeared to be flaring suggested an $\dot M$ increase. However, there has been some confusion in
the Cyg-like sources between dipping and flaring 
in that {\it reductions} in X-ray intensity in dipping (seen in hardness-intensity) when plotted in colour-colour 
look like a FB. In Cyg\th X-2 recent multi-wavelength observations show that intensity reductions were indeed
absorption dips unrelated to flaring (Ba\l uci\'nska-Church et al. 2011, 2012).
In GX\th 340+0 and \hbox{GX\th 5-1}, a 4th track has been seen at the end of the FB with decreasing intensity
suggesting a dipping-flaring link. However, spectral fitting has shown that this was not absorption 
but a reduction in the Comptonized emission level (Church et al. 2010). Thus the original evidence for the 
monotonic increase of $\dot M$ becomes questionable.

Support for the monotonic increase of $\dot M$ was found in the timing properties of the sources 
with many properties correlating with Z-track position, i.e. arc length $S$ along the track. 
However, in recent years authors have found several properties that do not correlate
(Wijnands et al. 1996; Homan et al. 2002; Casella et al. 2000; Dieters \& van der Klis 2000) 
so that $S$ is not a measure of $\dot M$.


In our model for the Cyg-like sources, it is proposed that $\dot M$ increases between 
the soft and hard apex. In the present work  we will test this proposal in the case of the Sco\th X-1
like sources, and find that this is also the case.

\subsection{Approach of the present work}

In this work, we test the hypothesis that the Extended ADC model can describe the 
Sco\th X-1 like sources, provide a physically reasonable explanation of them
and of the differences between these and the Cyg-like sources. 

It is applied in the form {\sc bb + cpl} where {\sc bb} is simple blackbody emission 
of the neutron star and {\sc cpl} is a cut-off power law representing Comptonization in an extended corona. 
As discussed in detail in Sect. 3.3, the model was applied with a fixed power law photon index of 1.7
and a free column density (except for Sco\th X-1). 
The {\sc bb + cpl} spectral form is appropriate to Comptonization of seed photons from the disk below the ADC which
at large radial positions have low energies ($kT$ $\sim$ 0.1 keV or less); see Church \& Ba\l uci\'nska-Church 
(2004) for a full discussion. A simple blackbody model is used and we have not previously found any evidence for 
departure from this form. Theoretical treatments of electron scattering in the neutron star atmosphere 
require measured blackbody temperatures to be corrected. Observational evidence 
for a modified blackbody is somewhat limited
as discussed in Ba\l uci\'nska-Church et al. (2001) who show that whether modification is important 
depends critically on the electron density which is poorly known. However, we show in Sect. 3.4 
that this correction combines with the relativistic correction to have a relatively small effect
on the blackbody temperature and radius.

\subsection{The sources: Sco\th X-1, GX\th 349+2 and GX\th 17+2}

Sco X-1 is the brightest persistent X-ray source in the sky (Giacconi et al. 1962) at a distance of 2.8$\pm $0.3 kpc 
(Bradshaw et al. 1999) and has been studied intensively since the early 1960s, including multi-wavelength 
campaigns (Canizares et al. 1975; Hertz et al. 1992). It has an orbital period of 18.9 hr (Cowley \& 
Crampton 1975) and a M type companion of mass 0.4 $M_{\sun}$ (Steeghs \& Casares 2002). Spectral 
modelling of Sco\th X-1 (e.g. White et al. 1985) showed that the spectra could be fitted by a blackbody 
plus Comptonization model. The third branch, the horizontal branch, was discovered by Hasinger et al. (1989) 
using {\it Exosat}. The first kHz QPO was discovered in this source by van der Klis et al. (1996).

GX\th 349+2 is relatively little studied but has a similar period of 21.85$\pm 0.4$ hr (Wachter \& Margon 1996). 
It has not been seen in the HB. A distance of 9.2 kpc was found by Penninx (1989) and 
van Paradijs \& White (1995) (see also Grimm et al. 2002). Zhang et al. (1998) discovered kHz QPO on the 
upper normal branch at 712 and 978 Hz. Iaria et al. (2004) found emission features probably
L-shell emission of Fe XXIV, Ly$\,{\alpha}$ S XVI and Fe XXV.

GX\th 17+2 appeared as a source with two branches (the NB and HB) in hardness-intensity (Schulz et al. 1989).
A strong FB was seen by Hasinger \& van der Klis (1989). The companion was difficult to identify (Deutsch et al. 1999; 
Callanan et al. 2002).
Investigations of the variation around the
Z-track of low frequency QPO were carried out by Langmeier et al. (1990), Kuulkers et al. (1997) and Homan
et al. (2002). Kilohertz QPO were discovered by Wijnands et al. (1997).
It is one of the two Z-track sources with occasional X-ray bursts: GX\th 17+2 and Cyg\th X-2 (Kahn \&
Grindlay 1984; Tawara et al. 1984; Sztajno et al. 1986; Kuulkers et al. 1997, 2002).
Migliari et al. (2007) from simultaneous radio and X-ray observations show the switch-on of the jet on the 
normal branch. Distances of about 7.5 kpc were found by several authors (e.g. Ebisuzaki et al. 1984); 
distances used in the analysis following are given in Table 6 and further discussed in Sect. 4.5.

\section{Observations and analysis}

All three sources have been observed on multiple occasions with {\it RXTE}. We examined the data in the HEASARC
archive remotely producing hardness-intensity plots and selected the best quality data, i.e. data which spanned 
as much as possible of a full Z-track. In the Sco-like sources most observations contain flaring, but few also 
have normal and horizontal branch data. In the case of GX\th 17+2, a full Z-track was available. It is also important 
not to use data in which sideways or secular movement of the Z-track takes place as we are interested in changes 
taking place around a single Z-track. The observations chosen are shown in Table 1.

\tabcolsep 1.0 mm
\begin{table}[!h]
\begin{center}
\caption{The {\it Rossi-XTE Observations.}}
\begin{minipage}{84mm}
\begin{tabular}{lrlrr}
\hline \hline
source               &obsid      &date      &length  & calib\\ 
                                          &&& (ksec) & epoch\\
\noalign{\smallskip\hrule\smallskip}
GX\th 349+2          &P30042    &1998 Jan 9-10  &170  &3\\
Sco\th X-1: No flare &P40706    &1999 Jun 10-13 &270  &4\\
Sco\th X-1: Flare    &P30036    &1998 Jan 7-8  &120  &3\\
GX\th 17+2           &P20053    &1997 Apr 1-4  &272  &3\\
\noalign{\smallskip}\hline
\end{tabular}\\
\end{minipage}
\end{center}
\end{table}


We previously published results on Sco\th X-1 for a typical observation dominated by strong flaring (P30036) made in 1998, 
Jan 7-8 spanning 120 ksec (Barnard et al. 2003). A major part of the present work relates to the effects of strong 
flaring seen in the Sco\th X-1 like sources, and to investigate this we searched the HEASARC archive for Sco\th X-1 
data in which there was substantially less flaring, if such data could be found. In the observation chosen for analysis 
(obsid P40706) there was a large gap of at least 150 ksec from any previous major flare, and we compare 
results for this with a reanalysis of the 1998 observation using the same improved calibration data. This observation
was also analysed by Bradshaw et al. (2003) using a different spectral approach.

In observations of all three sources, data from both the Proportional Counter Array (PCA, Jahoda et al. 1996) and 
the high-energy X-ray timing experiment (HEXTE) were used. The PCA was in Standard2 mode with 16$\,$s resolution and 
examination of the housekeeping data revealed the number of the xenon proportional counter units (PCUs) that were 
reliably on during each observation. In GX\th 349+2 and GX\th 17+2 all 5 PCUs were reliably on and data used from 
all of these with both left and right detectors. In both observations of Sco\th X-1, data from a single PCU and a 
single detector were used as the count rate was already high, and similar to that in the other
sources because the source distance is $\sim $3 times less. Standard screening criteria based on the housekeeping data 
were applied to select data having an offset between the source and telescope pointing of less than 0.02$^o$ and 
elevation above the Earth's limb greater than 10$\degmark$. Data were extracted from the top layer of the detector. 
Analysis was carried out using the {\sc ftools 6.9} package. The P40076 observation of Sco \th X-1 was made in Epoch 4 of 
the mission (1999 March 22 -- 2000 May 13) relevant to channel-energy conversion and also to calculation of background 
data from appropriate models. The other sources were observed in Epoch 3 (1996 April 15 -- 1999 March 22). Total PCA 
lightcurves were extracted to correspond as closely as possible to the band 2.0 -- 18.5 keV for all observations by 
selecting appropriate channel ranges.
Lightcurves were generated and 
background files for each PCA data file were produced using the facility {\sc pcabackest}, applying the latest ``bright'' 
background models appropriate to Epoch 3 or Epoch 4 for background subtraction of the lightcurves. Deadtime 
correction was carried out on both source and background files prior to background subtraction and lightcurves 
with 64 s binning are shown in Fig. 1.

\begin{figure*}[!ht]
\begin{center}                                                       
\includegraphics[width=46mm,height=176mm,angle=270]{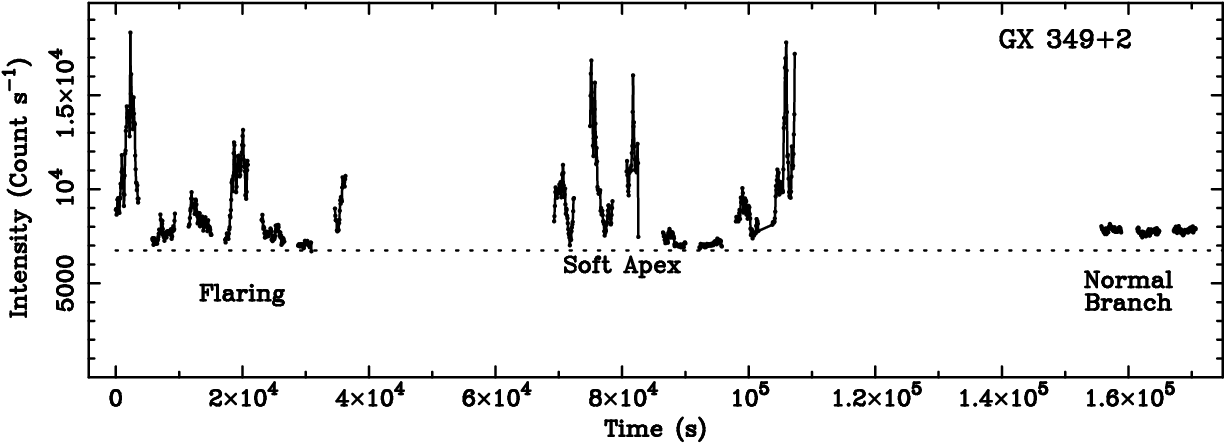}      
\includegraphics[width=46mm,height=176mm,angle=270]{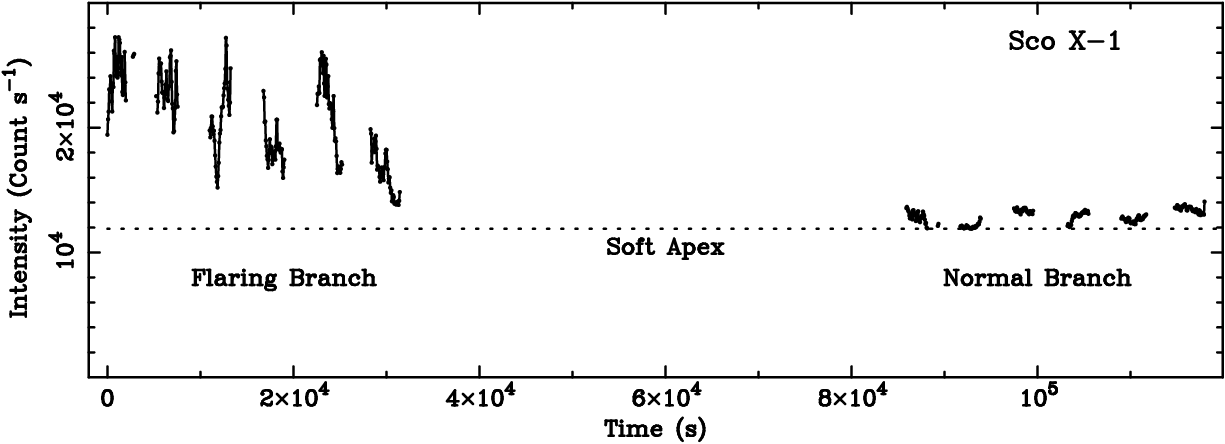}      
\includegraphics[width=46mm,height=176mm,angle=270]{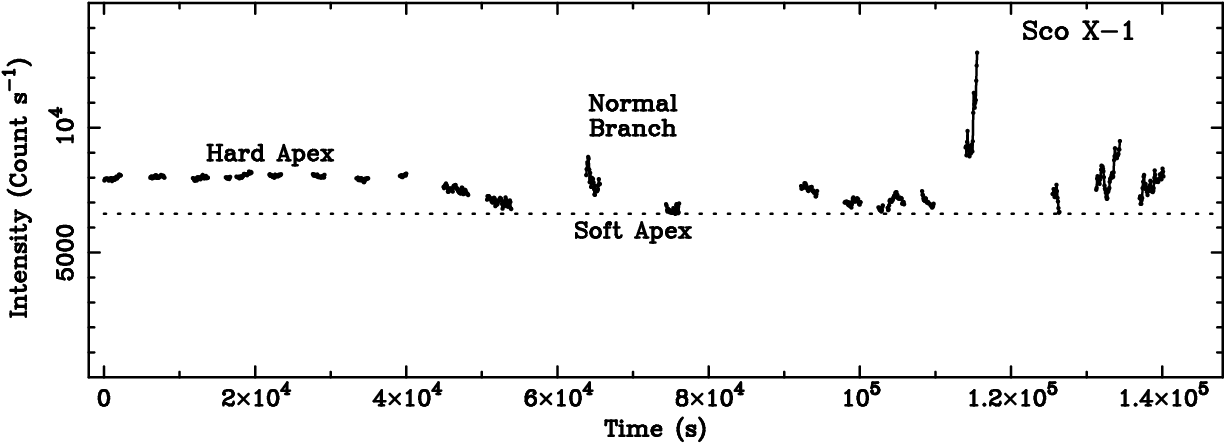}  
\includegraphics[width=46mm,height=176mm,angle=270]{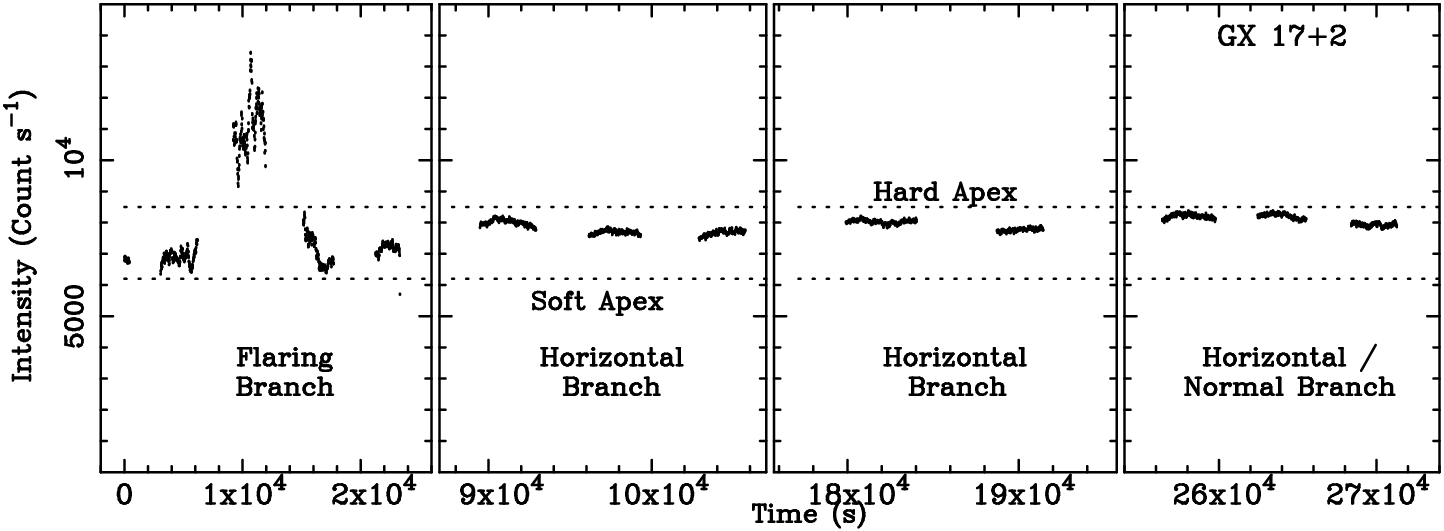}              
\caption{Lightcurves from the {\it Rossi-XTE} PCA, background-subtracted and deadtime-corrected 
of the 3 sources with 64 s binning. Upper panel: GX\th 349+2; upper middle panel: typical Sco\th X-1
data with strong flaring; lower middle panel: Sco\th X-1 in an exceptionally flare-free state;
lower panel: GX\th 17+2; the gaps between the 4 sub-observations are removed for clarity.}
\label{}
\end{center}
\end{figure*}
%
In each source an identification of the subsections of data in the lightcurve with branches of
the Z-track was made by selecting each of these sections with the ftool {\sc maketime} and making a
hardness-intensity plot of the selected data to find where it lay on the overall Z-track. This allowed
the sections of the lightcurves to be labelled as HB, NB or FB.

Lightcurves were also made in sub-bands of the above range, corresponding to energies 2.0 -- 4.1 keV,
4.1 -- 7.3 and 7.3 -- 18.5 keV for Epoch 3 data, and as close as these to possible in Epoch 4 by careful
choice of channels. A hardness ratio was defined as the ratio of the intensity in the band 7.3 -- 18.5 keV 
to that in the 4.1 -- 7.3 keV band. The resulting Z-tracks of hardness versus total intensity are shown 
in Fig. 2. The rebinning of the lightcurves from 16 s to 64 s was done to produce the 
relatively narrow width of the Z-track, as previous work has shown it is important to select data 
along the centre of the Z-track so spectral changes reflect changes along the track and not 
perpendicular to it (Barnard et al. 2003). Superimposed on the Z-tracks are boxes showing the selection
of spectral data (below). 

HEXTE data were also extracted as lightcurves and spectra from a single cluster of the HEXTE instrument 
using {\sc hxtlcurve} which also provided background files, and deadtime correction was made. 
This allowed simultaneous fitting of PCA and HEXTE spectra, the wider energy band resulting in spectral 
fitting being better constrained than with the PCA alone.

\section{Results}

\subsection{Lightcurves}

Figure 1 shows the lightcurves of the three sources, including the two observations of Sco\th X-1. In 
GX\th 349+2 and the 1998 observation of Sco\th X-1 (upper middle panel), the prevalence of strong flaring 
can be seen. By remotely scanning all of the archival {\it RXTE} data on Sco\th X-1 it is clear that
the source is essentially flaring most of the time. The 1999 observation of Sco\th X-1 (lower middle panel)
is highly unusual. It was chosen from all of the observations of Sco\th X-1 made during the 15 year life 
of the mission to be one of the few having a long period without flaring. Only towards the end of the observation,
is a strong flare seen with a factor of two increase of intensity. The 
lightcurve is shown for clarity without additional data spanning 120 ksec before the data shown, during
which only a weak flare was seen. The data analysed between 0 and 110 ksec appears free from flaring
and even if flaring occurred immediately before the additional data, the data analysed are 
$\sim$150 ksec or more later. This observation of Sco\th X-1 was made with an offset pointing (24\arcmin $\,$) 
as is common with this very bright source, so the count rate is reduced. In the case of GX\th 17+2, relatively 
little flaring can be seen, typical of the source, and this will be demonstrated and discussed later, it 
appearing that the source is transitional between the Sco-like and the Cyg-like sources.

\subsection{Z-tracks}

In Fig. 2 we show the Z-tracks of the 4 observations as hardness-intensity diagrams; the intensities are 
both background-subtracted and deadtime corrected. The observation of Sco\th X-1 with strong flaring is shown
as the upper central panel while that with little flaring is in the lower central panel. It can be 
seen that in GX\th 349+2 and the flaring Sco\th X-1 observation the horizontal branch is missing as is typical 
of the Sco-like sources. In the reduced flaring Sco\th X-1 observation, much more of the Z-track is present but it
is difficult to locate the hard apex exactly as the track is vertical in this region. In 
GX\th 17+2 the Z-track is complete, and flaring less persistent which as discussed later is a characteristic of 
this source which is more similar to the Cyg\th X-2 like group. PCA spectra were extracted in each case 
at several positions along the Z-track within hardness-intensity boxes by use of a 
good time interval (GTI) for each selection. The intensity ranges were typically 100 count s$^{-1}$
wide and hardness ranges such as 0.480 - 0.485 were used. It was checked that the selections
were correct by overlaying the selected data on the full Z-track. A typical PCA spectrum at a mid-range intensity
contains $\sim$10 Million counts equivalent to $\sim$1500 s of data so that high quality spectral fitting 
could be carried out. 
\begin{figure}[!ht]
\begin{center}                                                                     
\includegraphics[width=46mm,height=62mm,angle=270]{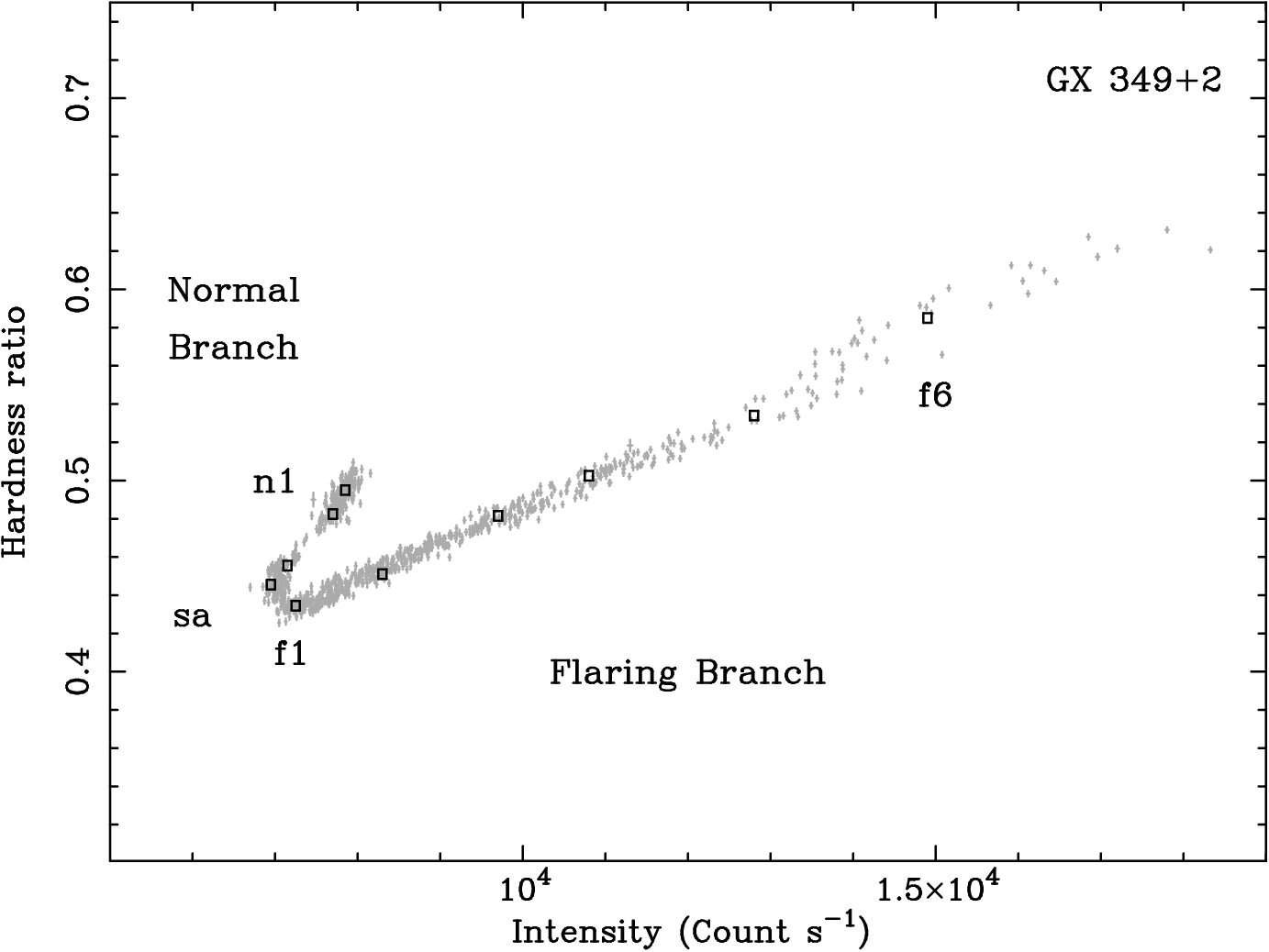}                     
\hskip 2mm
\includegraphics[width=46mm,height=62mm,angle=270]{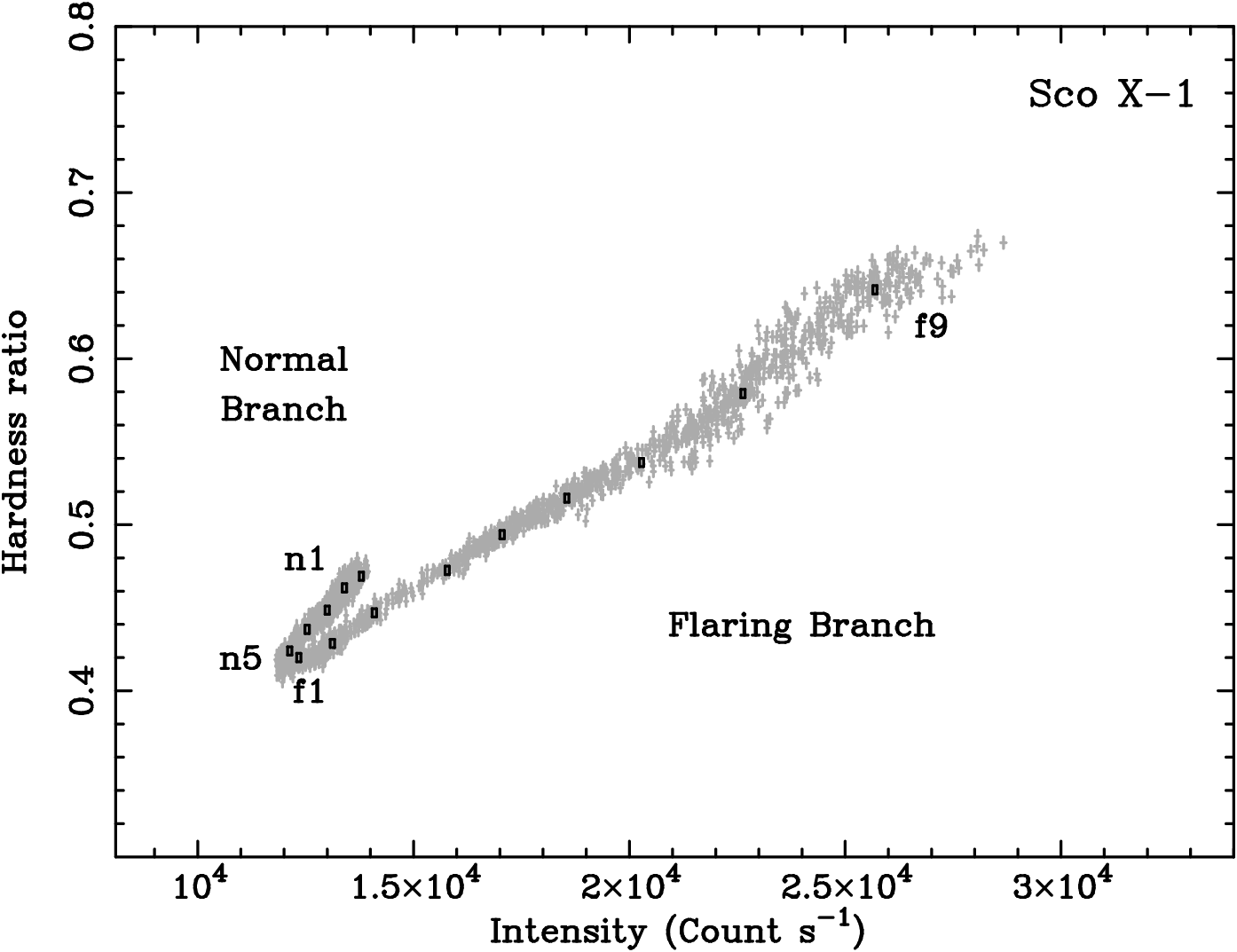}                     
\hskip2mm
\includegraphics[width=46mm,height=62mm,angle=270]{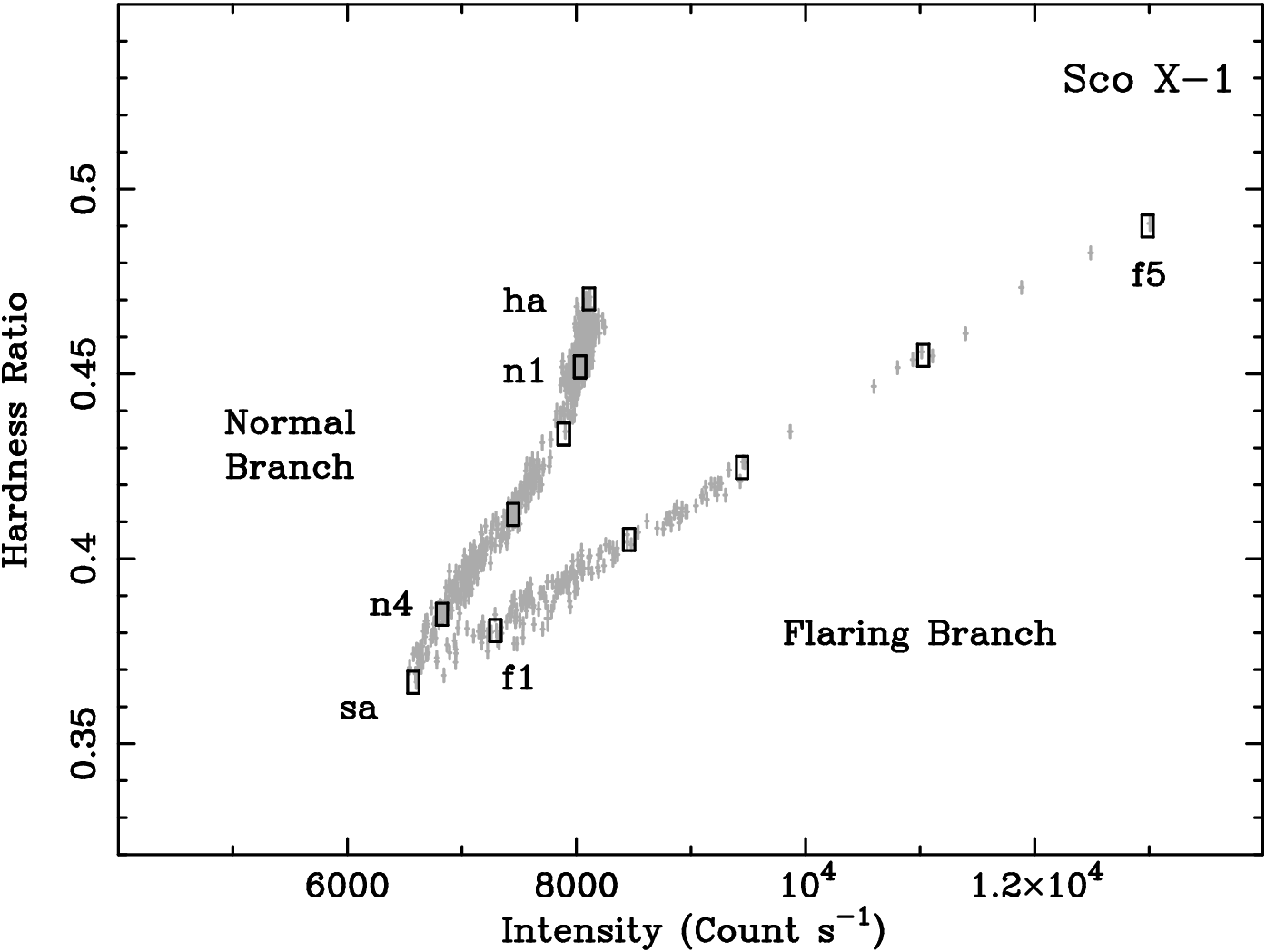}   
\hskip2mm
\includegraphics[width=46mm,height=62mm,angle=270]{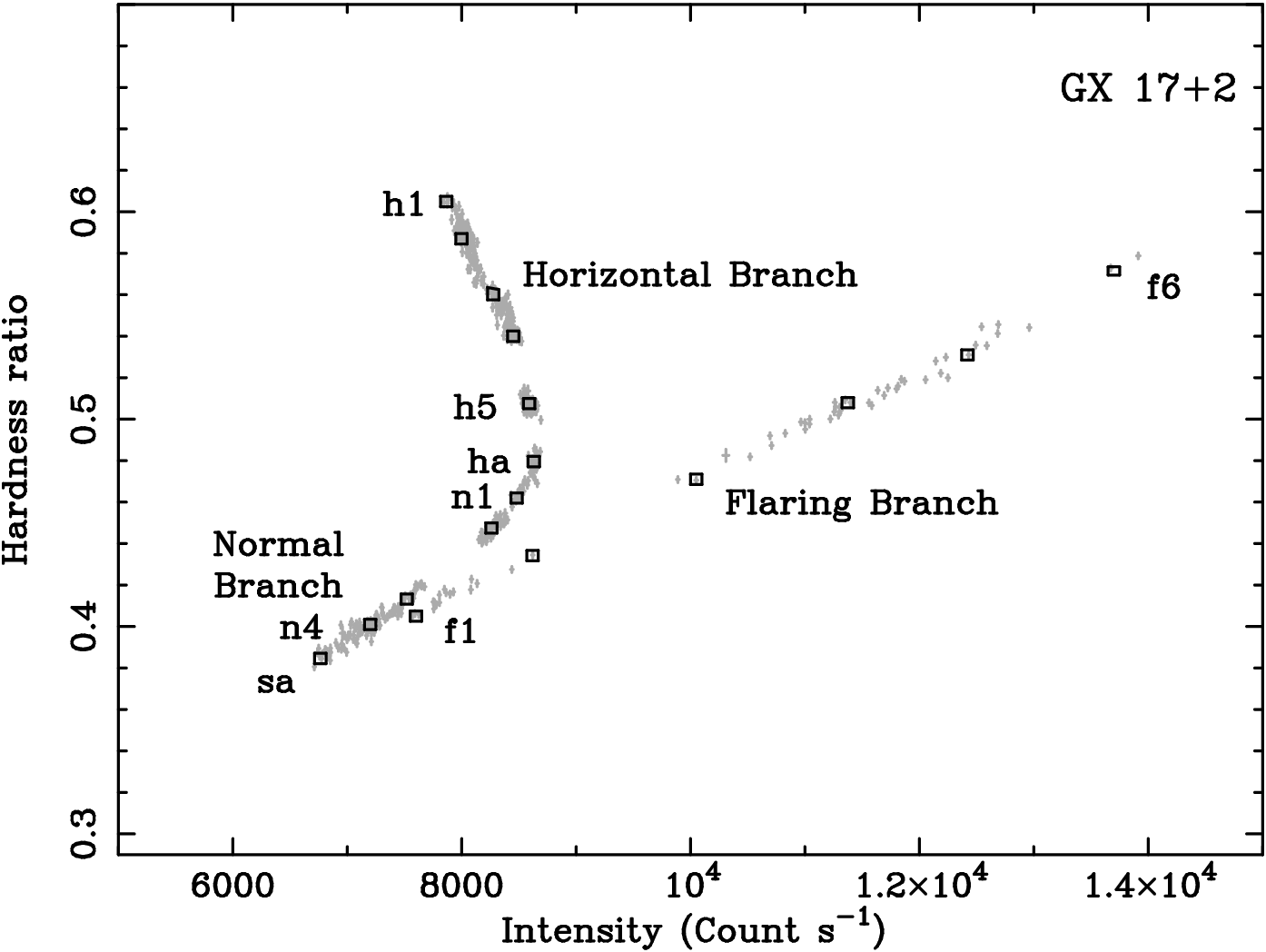}                     
\caption{Z-track of the observation derived from background subtracted and deadtime corrected
lightcurves with 64$\,$s binning. The boxes show the ranges of intensity and hardness ratio used
for the selection of PCA and HEXTE spectra.}
\label{}
\end{center}
\end{figure}
Background spectra were produced for each selection and source and background spectra deadtime-corrected
using local software {\sc pcadead}. Pulse pileup had negligible effect on the spectra in general and was
not carried out except in the case of the Sco\th X-1 observations. It was
not necessary to re-group channels to a minimum count to allow use of the $\chi^2$ statistic as 
the count in primitive channels was already high. A systematic error of 1.0\% was added to each channel as
is usual for PCA data. Response files appropriate to each observation were generated using {\sc pcarsp}
and the offset pointing in the P40706 observation of Sco\th X-1 taken into account.

HEXTE source and background spectra were extracted using the same GTI files as for the PCA and deadtime corrected
using {\sc hxtlcurv}. A standard auxiliary response file (arf) and response matrix file (rmf) were used. 
The rmf file was rebinned to match the actual number of HEXTE channels using {\sc rddescr}
and {\sc rbnrmf}. The source + background spectra were compared with the background spectra for both the
PCA and HEXTE, and spectral fitting carried out only up to the energy where these became equal, typically
22 keV in the PCA and 40 keV in HEXTE.

\subsection{Spectral analysis}

The motivation of this work is to test the hypothesis that the Extended ADC model 
may provide good fits to the spectra of the Sco\th X-1 like Z-track sources, and the model was applied 
in the form {\sc abs}$\ast ${(\sc bb + cpl)} (Sect. 1) where {\sc abs} is the {\sc wabs} absorption 
term in {\it Xspec}.

PCA and HEXTE spectra were fitted simultaneously allowing a variable normalization
between the instruments with parameter values chained between the PCA and HEXTE.
The lower energy limit of the PCA, usually set in spectral fitting at $\sim$3 keV, does not
allow accurate determination of low values of the column density. However, in GX\th 349+2 and
GX\th 17+2, $N_H$ values larger than $3\times 10^{22}$ atom cm$^{-2}$ were obtained by spectral fitting
and there is no problem in measuring such values. In the case of Sco\th X-1, the column density is much smaller
and was fixed at $0.3\times 10^{22}$ atom cm$^{-2}$ which was found to give good quality fits, although reducing
the value further made little difference to the fitting.
In GX\th 349+2, $N_{\rm H}$ at the soft apex is about 10 times larger than the Galactic value of 
$\sim 0.6\times 10^{22}$ atom cm$^{-2}$  (Dickey \& Lockman 1990; Kalberla et al. 2005), and in GX\th 17+2, 
$N_{\rm H}$ is several times larger than the Galactic $\sim 0.9\times 10^{22}$ 
atom cm$^{-2}$. This was also found in our previous work on the Cygnus\th X-2 like sources suggesting
that there is intrinsic absorption. Such higher values have been found previously in Z-track sources 
(Christian \& Swank 1997).

It is well-known that the spectra of the Z-track sources have a low Comptonization cut-off energy
of a few keV, 
which restricts the energy range available for determination of
the power law photon index $\Gamma$. We have taken the approach previously adopted (Paper I, II and III)
of fixing $\Gamma$ at 1.7, which seems to be optimum for fitting.
This procedure gave good quality spectral fits and if $\Gamma$ was freed, 
the value stayed close to 1.7. Extensive testing has shown that the value of power law index did not
affect the pattern of  variation of spectral parameters along the Z-track. In all three sources, initial fitting
revealed excess emission in the residuals and an Fe line was added to the model. In
GX\th 349+2 an additional edge feature at an energy of about 10 keV was clearly present and was added to
the model. An edge feature at $\sim$9 keV was seen previously by di Salvo
et al. (2001). Final fitting results are shown in Tables 2 - 5 for the four observations analysed and
representative spectral fits are shown in \hbox{Fig. 3.}

\subsection{Variation of the neutron star emission around the Z-track}

We now present results for the three \hbox{Sco\th X-1} like sources, using at this stage only
the observation that is typical of Sco\th X-1 with strong flaring. We concentrate on the properties
of the continuum emission. In Fig. 4 we show the variation of
the blackbody temperature $kT_{\rm BB}$ and blackbody radius $R_{\rm BB}$ around the Z-track as a
function of the total 1 - 30 keV luminosity (the sum of the blackbody, Comptonized emission and any
line present). Results are presented as derived from spectral fitting without making a correction
for general relativity as we wish to make comparisons between the Sco-like and the Cyg-like sources
as directly as possible.  At the soft apex between NB and FB, $kT_{\rm BB}$ is lower in all sources
than on the NB, increasing
on the normal branch and in the case of GX\th 17+2 which has a horizontal branch, also increasing on
that. In Sco\th X-1 and GX\th 349+2, $kT_{\rm BB}$ remains always greater than 2 keV, staying
about constant in flaring, while in GX\th 17+2 this temperature is reached at flare peaks and on the upper NB.
\begin{figure*}[!ht]                                                
\begin{center}
\includegraphics[width=44mm,height=70mm,angle=270.]{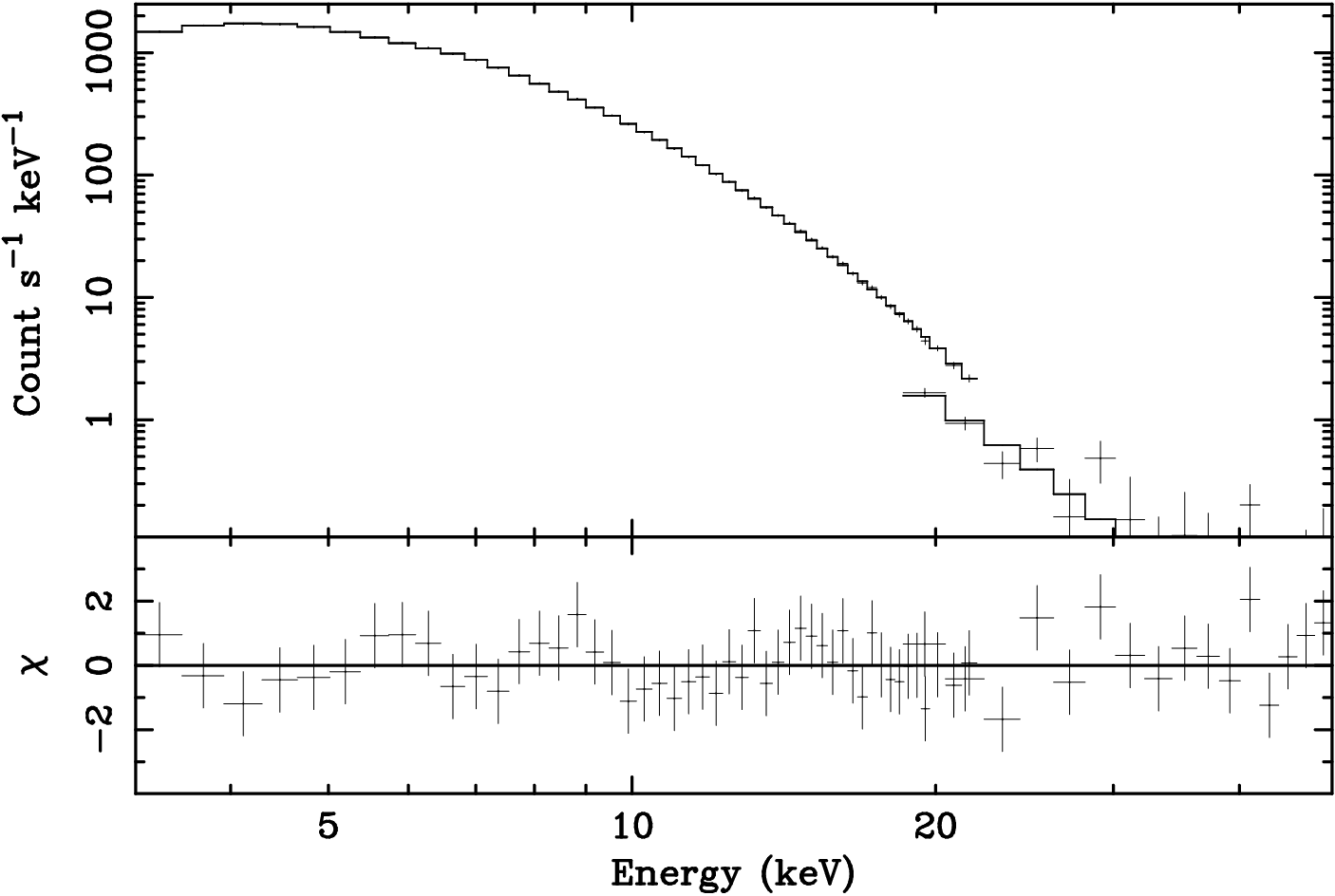}    
\includegraphics[width=44mm,height=70mm,angle=270.]{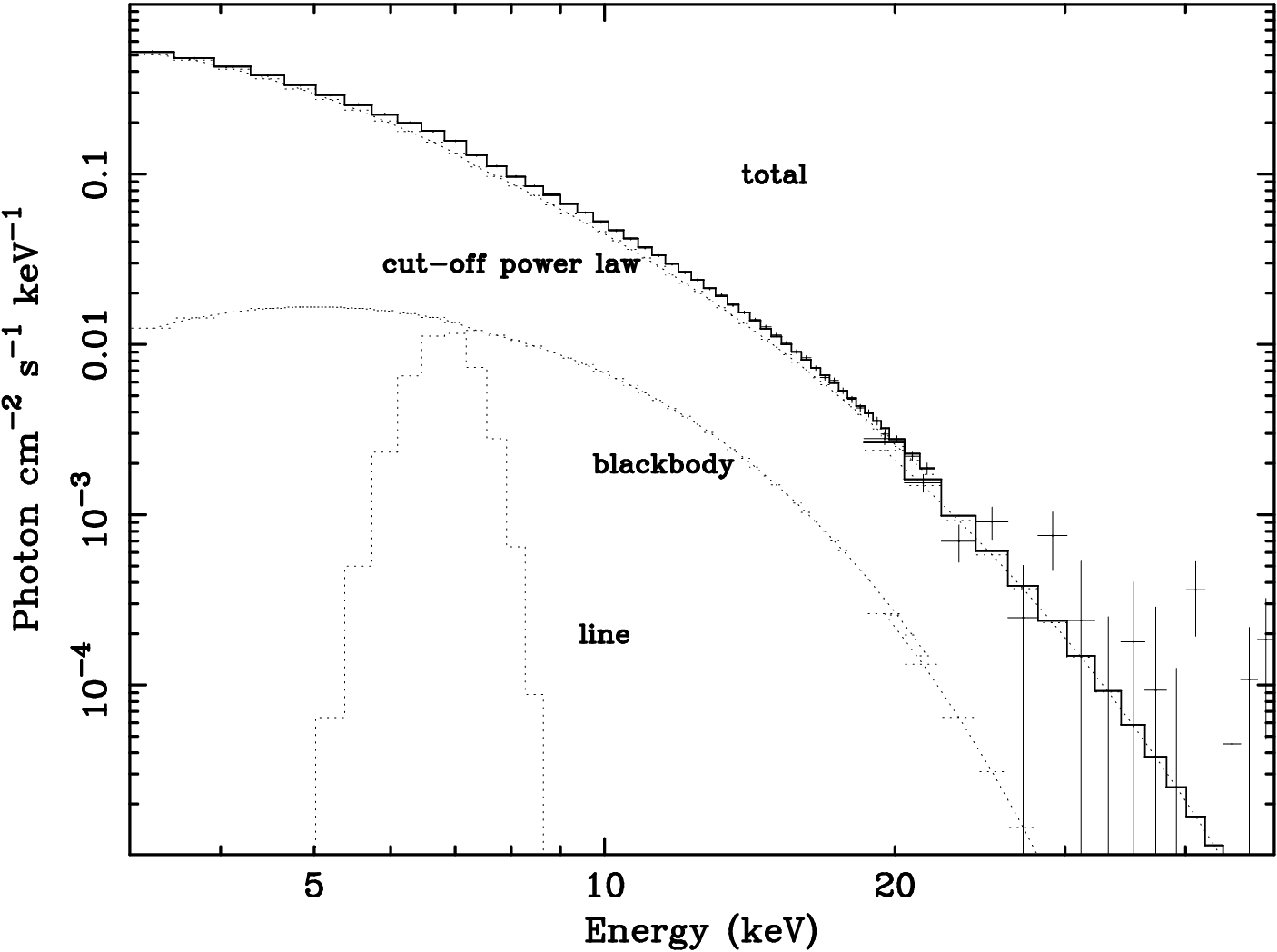}    
\includegraphics[width=44mm,height=70mm,angle=270.]{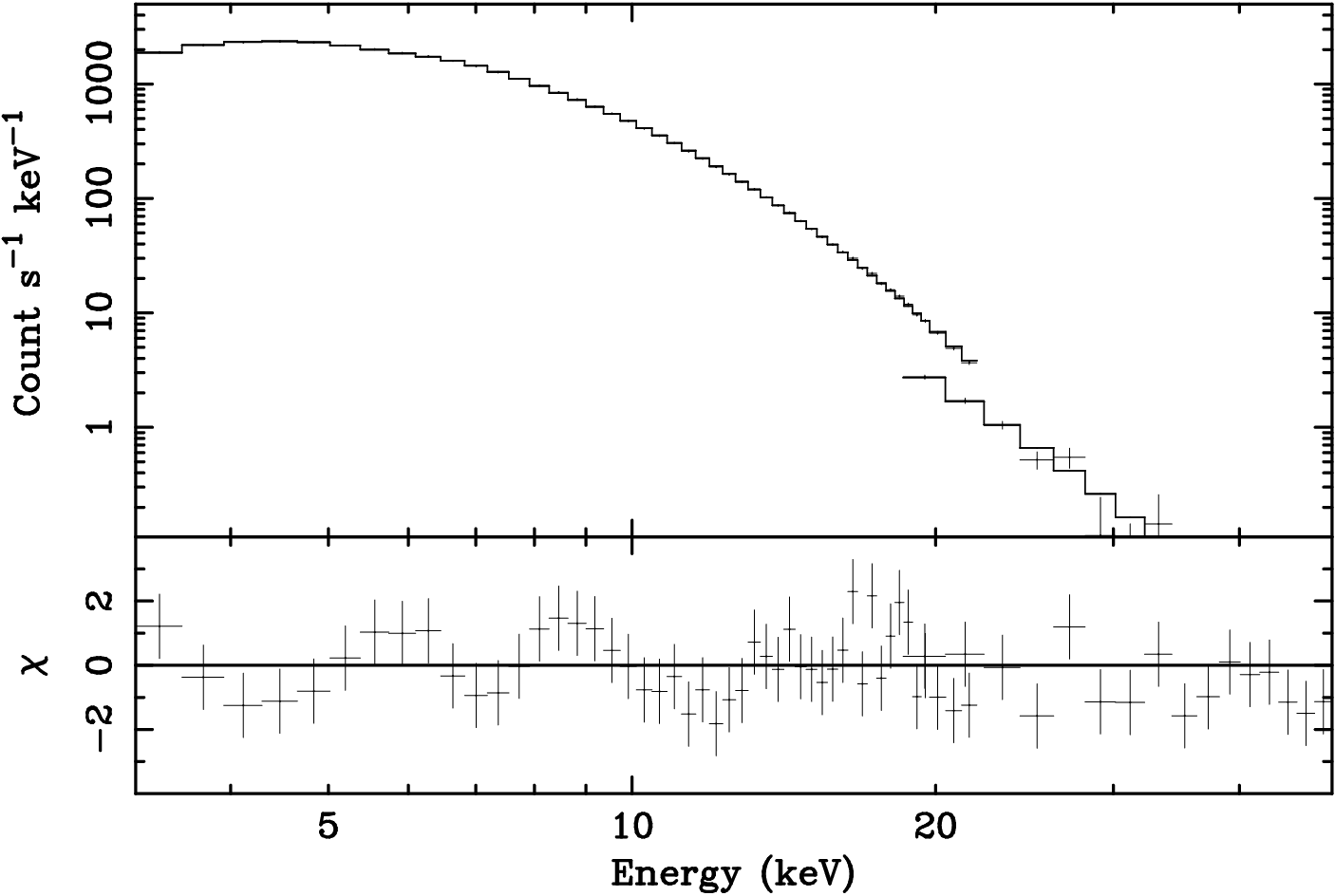}    
\includegraphics[width=44mm,height=70mm,angle=270.]{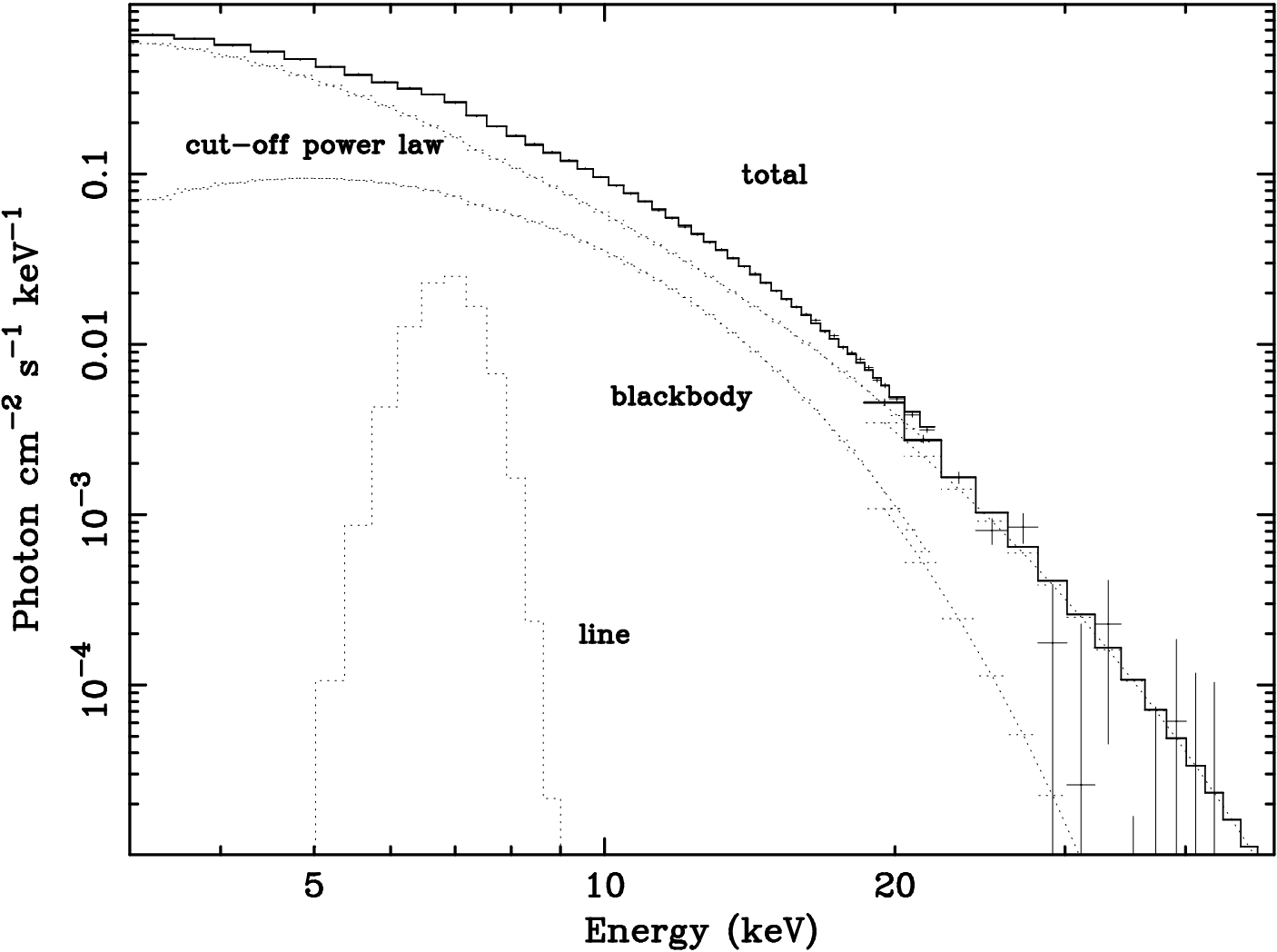}    
\caption{Best fit spectra for GX\th 17+2 in the NB (top; spectrum n2) and FB (bottom; spectrum f5),
showing the folded data (left) with residuals and unfolded data (right). The error bars are generally
too small to be visible. Model components are labelled in the unfolded spectra.}
\label{}
\end{center}
\end{figure*}
\vskip 4 mm
\tabcolsep 0.5 mm                                                                       
\begin{table*}[!h]
\begin{center}
\caption{GX\th 349+2 spectral fitting results}
\begin{tabular}{llrrrlrrrrlr}
\hline \hline
$\;\;$spectrum&$N_{\rm H}$&$kT_{\rm BB}$&$norm_{\rm BB}$&$R_{\rm BB}$&$E_{\rm CO}$&$norm_{\rm ADC}$&$E_{\rm line}$&{\it EW}&$E_{\rm edge}$&{\hskip 8mm}$\tau$&$\chi^2$/d.o.f.\\
&& (keV) && (km) & (keV) && (keV) & (eV) & (keV)\\
\noalign{\smallskip\hrule\smallskip}
Normal &{\hskip -4mm}Branch\\
n1 &3.6$\pm 0.8$  &2.35$\pm 0.09$  &6.48$\pm 0.87$ &3.7$\pm 0.4$ &5.2$\pm 0.6$  &9.99$\pm 1.2$  &6.57$\pm 0.34$  &144  &10.06$\pm 0.81$  &0.04$\pm 0.03$ &20/54\\
n2 &3.5$\pm 0.8$  &2.27$\pm 0.13$  &5.47$\pm 0.82$ &3.7$\pm 0.5$ &5.5$\pm 0.7$  &9.52$\pm 1.2$  &6.53$\pm 0.33$  &153  &10.09$\pm 1.20$  &0.04$\pm 0.03$ &26/54\\
n3 &3.3$\pm 0.8$  &2.19$\pm 0.12$  &4.44$\pm 0.61$ &3.6$\pm 0.5$ &5.4$\pm 0.8$  &8.94$\pm 1.4$  &6.43$\pm 0.23$  &198   &9.91$\pm 0.46$  &0.06$\pm 0.03$ &34/54\\
\noalign{\smallskip}
Soft &Apex\\
sa &3.0$\pm 0.8$  &2.01$\pm 0.16$  &4.10$\pm 0.81$ &4.1$\pm 0.7$ &5.6$\pm 0.5$  &8.40$\pm 1.3$  &6.44$\pm 0.26$  &203   &9.55$\pm 0.83$  &0.05$\pm 0.04$ &38/54\\
\noalign{\smallskip}
Flaring &{\hskip -4mm}Branch \\
f1 &3.8$\pm 0.8$  &2.08$\pm 0.12$  &4.49$\pm 0.70$ &4.0$\pm 0.6$ &5.0$\pm 0.4$  &10.1$\pm 1.3$  &6.48$\pm 0.21$  &239   &9.84$\pm 0.46$  &0.07$\pm 0.03$&37/54\\
f2 &4.1$\pm 0.8$  &2.05$\pm 0.10$  &5.48$\pm 0.66$ &4.5$\pm 0.7$ &5.2$\pm 0.3$  &10.9$\pm 1.4$  &6.55$\pm 0.20$  &247  &10.03$\pm 0.47$  &0.07$\pm 0.03$&32/54\\
f3 &4.6$\pm 0.9$  &2.08$\pm 0.11$  &8.49$\pm 0.90$ &5.5$\pm 0.7$ &5.4$\pm 0.5$  &12.6$\pm 2.0$  &6.62$\pm 0.23$  &236  &10.22$\pm 0.53$  &0.08$\pm 0.03$&61/54\\
f4 &4.4$\pm 0.9$  &2.03$\pm 0.08$ &10.47$\pm 1.06$ &6.4$\pm 0.6$ &5.7$\pm 0.4$  &12.0$\pm 1.9$  &6.63$\pm 0.25$  &205  &10.32$\pm 0.45$  &0.08$\pm 0.03$&42/54\\
f5 &4.2$\pm 0.9$  &2.08$\pm 0.08$ &15.21$\pm 0.72$ &7.3$\pm 0.6$ &6.3$\pm 0.6$  &11.9$\pm 2.0$  &6.62$\pm 0.28$  &191  &10.24$\pm 0.48$  &0.08$\pm 0.03$&65/54\\
f6 &4.2$\pm 0.9$ &2.07$\pm 0.06$  &20.15$\pm 1.57$ &8.7$\pm 0.6$ &6.9$\pm 0.6$  &11.3$\pm 2.0$  &6.81$\pm 0.32$  &150  &10.62$\pm 0.43$  &0.08$\pm 0.03$&31/54\\
\noalign{\smallskip}\hline
\end{tabular}\\
\end{center}
\hskip 10 mm All Tables: column densities are in units of 10$^{22}$ atom cm$^{-2}$; the normalization of the blackbody is in units of\\
\vskip -4 mm \hskip 10 mm $\rm {10^{37}}$ erg s$^{-1}$ for a distance of 10 kpc, the cut-off power law normalization for the ADC emission is in units of\\
\vskip -4 mm \hskip 10 mm photon cm$^{-2}$ s$^{-1}$ keV$^{-1}$ at 1 keV.  90\% confidence uncertainties are shown.\\
\end{table*}
%
\tabcolsep 2.0 mm                                                                     
\begin{table*}[!h]
\begin{center}
\caption{Sco X-1 with strong flaring}
\begin{tabular}{lrrrrrrrrrr}
\hline \hline
\noalign{\smallskip}
$\;\;$spectrum &$kT_{\rm BB}$&$norm_{\rm BB}$&$R_{\rm BB}$&$E_{\rm CO}$&$norm_{\rm ADC}$&$E_{\rm line}$& $EW$&$\chi^2$/d.o.f.\\
&(keV)&&(km)&(keV)&&(keV)&(eV)\\
\noalign{\smallskip\hrule\smallskip}
Normal Branch\\
%
n1& 2.33$\pm 0.07$ &85.6$\pm 6.8$& 4.20$\pm 0.30$&5.35$\pm 0.37$& 171.3$\pm 5.8$ &6.39$\pm 0.24$ &257 &43/55\\
n2& 2.16$\pm 0.05$ &73.3$\pm 4.5$& 4.52$\pm 0.25$&5.71$\pm 0.08$&160.2$\pm 3.7$ &6.42$\pm 0.22$ &175 &36/55\\
n3& 2.17$\pm 0.06$ &65.8$\pm 4.3$& 4.25$\pm 0.26$&5.53$\pm 0.08$&161.8$\pm 3.8$& 6.40$\pm 0.22$ & 208 & 33/55\\
n4& 2.13$\pm 0.08$ &58.3$\pm 5.1$& 4.14$\pm 0.35$& 5.50$\pm 0.11$& 158.2$\pm 4.4$& 6.37$\pm 0.23$ &239 &32/55\\
\noalign{\smallskip}
Soft Apex\\
sa& 2.06$\pm 0.09$& 52.8$\pm 5.6$& 4.24$\pm 0.44$&5.44$\pm 0.13$ & 155.7$\pm 4.5$& 6.34$\pm 0.16$ & 236 & 42/55\\
\noalign{\smallskip}
Flaring Branch\\
f1& 2.02$\pm 0.08$& 59.0$\pm 5.3$& 4.62$\pm 0.43$& 5.26$\pm 0.14$ & 160.0$\pm 4.8$ & 6.33$\pm 0.20$ &276  & 47/55\\
f2& 1.97$\pm 0.07$& 78.2$\pm 6.1$& 5.62$\pm 0.45$& 5.22$\pm 0.15$ & 164.1$\pm 5.3$ & 6.33$\pm 0.19$ &259  & 49/55\\
f3& 2.00$\pm 0.07$&101.0$\pm 7.1$& 6.18$\pm 0.45$& 5.18$\pm 0.18$ & 170.1$\pm 6.2$ & 6.34$\pm 0.20$ &264  & 32/55\\
f4& 2.01$\pm 0.06$&132.3$\pm 9.0$& 6.98$\pm 0.50$& 5.23$\pm 0.21$ & 180.5$\pm 7.5$ & 6.42$\pm 0.20$ &254  & 45/55\\
f5& 2.03$\pm 0.04$&154.4$\pm 8.2$& 7.42$\pm 0.37$& 5.43$\pm 0.13$ & 184.2$\pm 6.0$ & 6.43$\pm 0.20$ &220  & 60/55 \\
f6& 2.10$\pm 0.03$&161.5$\pm 8.5$& 7.59$\pm 0.39$& 5.40$\pm 0.20$ & 192.6$\pm 4.7$ & 6.43$\pm 0.20$ &263  & 61/45\\
f7& 2.18$\pm 0.04$&208.3$\pm 12$ & 7.49$\pm 0.37$& 5.36$\pm 0.25$ & 214.4$\pm 8.5$ & 6.52$\pm 0.22$ &259  & 47/46\\
f8& 2.05$\pm 0.03$&271.3$\pm 11$ & 8.17$\pm 0.27$& 5.50$\pm 0.20$ & 210.2$\pm 6.7$ & 6.41$\pm 0.22$ &235  & 28/40 \\
f9& 1.99$\pm 0.03$&341.9$\pm 8.1$& 7.72$\pm 0.27$& 6.15$\pm 0.15$ & 214.8$\pm 8.0$ & 6.41$\pm 0.20$ &250  & 33/46\\
\noalign{\smallskip}\hline
\end{tabular}\\
\end{center}
\end{table*}
%
\vskip 12 mm
\tabcolsep 1.8 mm                                                                     
\begin{table*}[!h]
\begin{center}
\caption{Sco X-1 with reduced flaring}
\begin{tabular}{lrrrrrrrrrr}
\hline \hline
\noalign{\smallskip}
$\;\;$spectrum &$kT_{\rm BB}$&$norm_{\rm BB}$&$R_{\rm BB}$&$E_{\rm CO}$&$norm_{\rm ADC}$&$E_{\rm line}$& $EW$&$\chi^2$/d.o.f.\\
&(keV)&&(km)&(keV)&&(keV)&(eV)\\
\noalign{\smallskip\hrule\smallskip}
%
Hard Apex\\
ha &2.28$\pm$0.05   &106.9$\pm$5.3  &4.89$\pm$0.24   &6.30$\pm$0.19   &129.5$\pm$4.2  &6.29$\pm$0.29  &159   &25.8/42\\

\noalign{\smallskip}
Normal Branch\\
n1 &2.26$\pm$0.03   &101.1$\pm$3.5  &4.84$\pm$0.15   &5.98$\pm$0.08   &135.3$\pm$3.5  &6.25$\pm$0.24  &188   &36.3/50\\
n2 &2.23$\pm$0.08   &88.6$\pm$6.8   &4.66$\pm$0.37   &5.85$\pm$0.28   &137.9$\pm$5.6  &6.33$\pm$0.20  &235   &35.9/50\\
n3 &2.14$\pm$0.05   &74.1$\pm$4.9   &4.64$\pm$0.28   &5.68$\pm$0.13   &138.0$\pm$4.4  &6.30$\pm$0.22  &215   &22.4/37\\
n4 &2.01$\pm$0.07   &53.7$\pm$4.5   &4.45$\pm$0.36   &5.66$\pm$0.13   &131.8$\pm$4.4  &6.27$\pm$0.19  &241   &28.0/44\\

\noalign{\smallskip}
Soft Apex \\
sa &1.83$\pm$0.14  &51.9$\pm$7.0   &5.26$\pm$0.85   &5.62$\pm$0.27   &126.8$\pm$7.5  &6.20$\pm$0.18  &254   &19.0/41\\

\noalign{\smallskip}
Flaring Branch\\
f1 &1.85$\pm$0.07    &72.5$\pm$6.5  &6.10$\pm$0.56   &5.57$\pm$0.18   &134.8$\pm$6.3  &6.30$\pm$0.17  &270   &24.8/43\\
f2 &1.87$\pm$0.06   &116.9$\pm$8.0  &7.60$\pm$0.51   &5.63$\pm$0.20   &140.0$\pm$7.1  &6.30$\pm$0.19  &256   &41.9/41\\
f3 &1.90$\pm$0.05   &151.3$\pm$8.6  &8.39$\pm$0.50   &5.75$\pm$0.13   &144.7$\pm$4.2  &6.39$\pm$0.20  &235   &35.6/40\\
f4 &1.94$\pm$0.04   &203.0$\pm$9.7  &9.34$\pm$0.42   &5.93$\pm$0.18   &154.2$\pm$7.5  &6.39$\pm$0.22  &210   &37.6/44\\
f5 &2.05$\pm$0.05   &262.0$\pm$14.2 &9.46$\pm$0.52   &5.95$\pm$0.32   &175.2$\pm$10.1 &6.42$\pm$0.21  &235   &32.0/41\\
\noalign{\smallskip}\hline
\end{tabular}\\
\end{center}
\end{table*}
%
%
\tabcolsep 0.5 mm                                                                     
\begin{table*}[!ht]
\begin{center}
\caption{GX\th 17+2 spectral fitting results}
\begin{tabular}{lrrrrrrrrrr}
\hline \hline
\noalign{\smallskip}
$\;\;$spectrum&$N_{\rm H}$&$kT_{\rm BB}$&$norm_{\rm BB}$&$R_{\rm BB}$&$E_{\rm
CO}$&$norm_{\rm ADC}$&$E_{\rm line}$& $EW$&$\chi^2$/d.o.f.\\
&&(keV)&&(km)&(keV)&&(keV)&(eV)\\
\noalign{\smallskip\hrule\smallskip}
Horizontal Branch\\
h1&3.6$\pm$0.6   & 2.78$\pm 0.20$  &5.94$\pm 1.43$ & 2.09$\pm 0.39$ & 8.43$\pm{1.48}$ & 7.34$\pm{0.79}$ & 6.71$\pm{0.24}$ & 107 & 44/53 \\
h2&4.1$\pm {0.5}$ & 2.79$\pm 0.12$ &5.62$\pm 0.75$ & 2.01$\pm 0.23$  &7.81$\pm{0.72}$  &8.24$\pm{0.66}$ &6.79$\pm{0.25}$ & 96 &55/56\\
h3&4.3$\pm {0.5}$& 2.73$\pm 0.13$  &5.15$\pm 0.76$& 2.01$\pm 0.25$ &7.34$\pm{0.64}$ &9.18$\pm{0.74}$ &6.72$\pm{0.25}$& 98 &32/56\\
h4&4.6$\pm {0.5}$ & 2.68$\pm 0.12$ &4.84$\pm 0.66$& 2.01$\pm 0.23$ &6.34$\pm{0.51}$ &10.08$\pm{0.75}$ &6.76$\pm{0.24}$ & 93 &53/56\\
h5&5.2$\pm {0.5}$ & 2.60$\pm 0.14$ &3.95$\pm 0.66$& 1.94$\pm 0.25$ &6.34$\pm{0.53}$ &11.74$\pm{0.87}$ &6.88$\pm{0.26}$& 98 &35/56\\

\noalign{\smallskip}
Hard Apex\\
ha&5.6$\pm {0.3}$ & 2.49$\pm 0.16$  &3.44$\pm 0.63$& 1.98$\pm 0.30$  &5.85$\pm{0.37}$&13.16$\pm{1.00}$ &6.91$\pm{0.23}$&98 &44/56\\

\noalign{\smallskip}
Normal Branch \\
n1&5.5$\pm {0.5}$ & 2.37$\pm 0.19$ &2.82$\pm 0.59$& 1.98$\pm 0.39$ &5.76$\pm{0.37}$&13.27$\pm{1.07}$ &6.86$\pm{0.24}$ &98 &45/56\\
n2&5.5$\pm {0.6}$ & 2.15$\pm 0.32$ &1.97$\pm 0.56$& 2.01$\pm 0.67$  &5.82$\pm{0.29}$ &13.12$\pm{1.27}$  &6.86$\pm{0.28}$ &95 &39/53\\
n3&5.4$\pm{0.7}$ & 1.65$\pm 0.24$  & 1.05$\pm 0.63$& 2.48$\pm 0.9$&5.62$\pm{0.28}$ &12.52$\pm{0.83}$ & 6.79$\pm{0.20}$ & 98 & 45/56\\
n4&5.5$\pm 0.7$ & 1.58$\pm 0.21$ & 1.65$\pm 0.90$& 3.41$\pm 1.30$ & 5.36$\pm{0.22}$ & 12.25$\pm{1.18}$ & 6.79$\pm 0.12$ &183 &62/56\\

\noalign{\smallskip}
Soft Apex \\
sa&4.4$\pm$0.9 & 1.28$\pm 0.07$ & 3.68$\pm 1.66$& 7.76$\pm 1.90$ &5.73$\pm{0.40}$ & 9.22$\pm{1.81}$ & 6.71$\pm{0.16}$ & 164 & 52/53\\

\noalign{\smallskip}
Flaring Branch\\
f1&4.7$\pm 0.7$ & 1.44$\pm 0.08$ & 4.37$\pm 1.54$& 6.61$\pm 1.35$ & 5.83$\pm$0.37 &10.05$\pm$1.77 & 6.65$\pm$0.14 & 175  & 36/53\\
f2&5.0$\pm 0.5$ & 1.54$\pm 0.09$ & 5.34$\pm 0.84$& 6.46$\pm 0.88$ &6.13$\pm 0.22$ & 10.90$\pm{0.68}$ & 6.72$\pm{0.15}$ & 151 & 45/53\\
f3&5.2$\pm 0.9$ & 1.75$\pm 0.11$ &7.23$\pm 1.64$& 5.81$\pm 1.08$ & 6.30$\pm{0.56}$ & 12.14$\pm{1.19}$ & 6.64$\pm{0.18}$ &147 & 54/53\\
f4&6.0$\pm {0.6} $ & 1.98$\pm 0.08$ &8.67$\pm 0.82$& 4.96$\pm 0.43$ &6.12$\pm{0.26}$ &14.62$\pm{1.37}$  &6.82$\pm{0.17}$ &120 &61/56\\
f5&5.9$\pm 0.6$ & 2.05$\pm 0.07$ & 10.69$\pm 0.81$& 5.13$\pm 0.39$ & 6.36$\pm{0.28}$ &14.85$\pm{1.38}$ & 6.89$\pm{0.20}$ & 120 & 66/56 \\
f6&5.2$\pm {0.8}$ & 1.99$\pm 0.10$ &14.07$\pm 1.55$& 6.28$\pm 0.72$ &7.80$\pm{0.81}$ & 12.51$\pm{2.05}$ & 6.84$\pm{0.30}$ & 93 & 42/53\\
\noalign{\smallskip}\hline
\end{tabular}\\
\end{center}
\end{table*}
\clearpage

\begin{figure}[!t]                                                
\begin{center}
\includegraphics[width=84mm,height=84mm,angle=270]{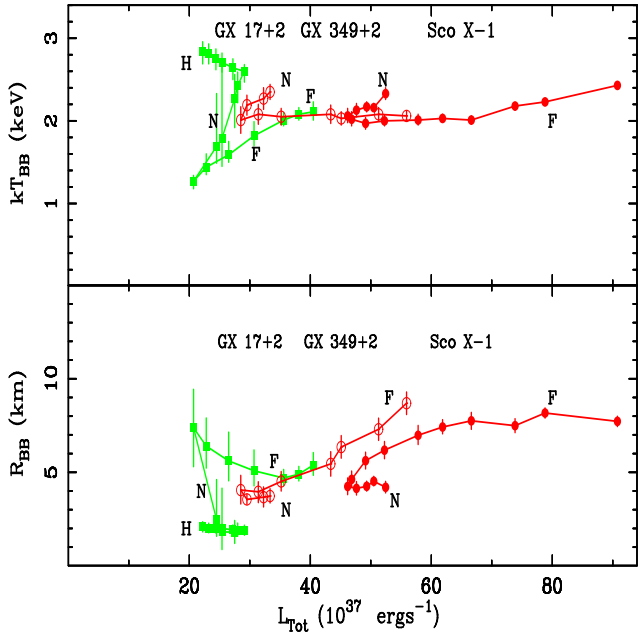}        
\caption{The neutron star blackbody emission in the Sco-like sources: GX\th 17+2: green filled squares,
GX\th 349+2: red open circles, Sco\th X-1: red filled circles. Upper panel: the blackbody temperature;
lower panel: the blackbody radius. Results are shown only for the observation of Sco\th X-1
in its normal state with strong flaring (obsid P30036). The branches are labelled H (horizontal),
N (normal) and F (flaring). In the electronic version of the paper, sources are shown with
individual colours.}
\label{}
\end{center}
\end{figure}


The blackbody radius $R_{\rm BB}$ shown in Fig. 4 is defined by
$L_{\rm BB}$ = 4$\pi \, R_{\rm BB}^2 \sigma T_{\rm BB}^4$ where $L_{\rm BB}$ is the blackbody
luminosity and $\sigma$ is Stefan's constant. It can be seen that the values at the soft apex
are depressed well below higher values found at other parts of the Z-track showing that
the whole neutron star is not emitting.

If we assume that both a relativistic correction and a colour correction for modification
of the blackbody by scattering in the neutron star atmosphere should be made (but see Sect. 1.4)
the corrected temperature $T_{\rm eff}$ = $T_{\rm c}\times (1+z)/f_{\rm c}$ where $T_{\rm c}$ is the
measured temperature. Using an appropriate  colour correction of $\sim$1.45 (Suleimanov et al. 2011),
and (1 + z) = 1.34, the overall correction would be only 9\%. The blackbody radius is corrected by
$f_{\rm c}^2/(1 + z)$ so that the real value would be 1.6 times larger than that measured.

\begin{figure}[!hb]                                                
\begin{center}
\includegraphics[width=84mm,height=84mm,angle=270]{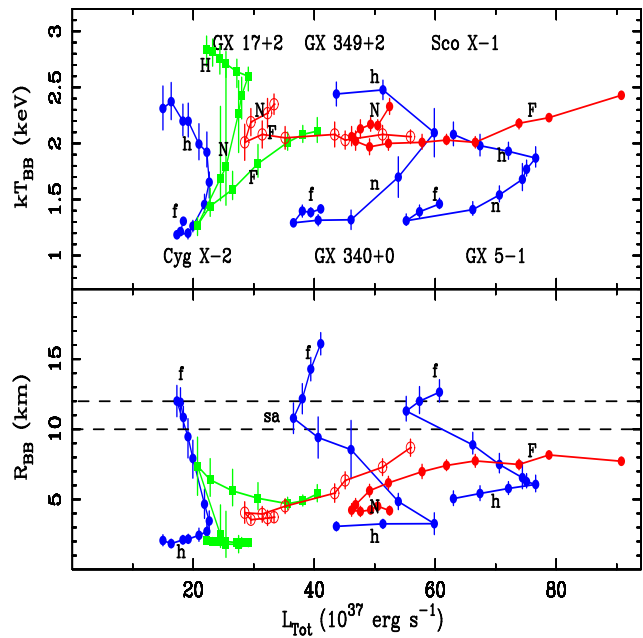}        
\caption{Comparison of the Sco\th X-1 like sources with the \hbox{Cyg\th X-2} like sources.
In all figures the Cyg-like sources will be shown as blue filled circles; GX\th 349+2 will
have red open circles and Sco\th X-1 red filled circles. The nominally Sco-like source GX\th 17+2
will be shown with green filled squares to emphasize its transitional nature with some characteristics
Cyg-like and some Sco-like. To help distinguish between the sub-groups of sources, Sco-like
sources have branch labels: H, N and F while Cyg-like sources use h, n and f.}
\label{}
\end{center}
\end{figure}

In Fig. 5 the results for the Sco-like sources are compared with those previously obtained 
for the Cyg-like sources. In all cases appropriate source distances from the literature as 
shown in Table 6 are used to determine luminosities.
In the Sco-like sources, $kT_{\rm BB}$ increases on the NB and HB (when present)
as was previously found in the three Cygnus\th X-2 like sources: Cygnus\th X-2, GX\th 340+0
and GX\th 5-1 (Papers I, II and III). However, the major difference between the Sco-like
and Cyg-like sources is clearly the substantially higher $kT_{\rm BB}$ at all
parts of the Z-track in the Sco-like sources. We will suggest that this has a major effect
on the inner disk. GX\th 17+2 appears transitional between the two types,
having a lower $kT_{\rm BB}$ at the soft apex but also having the strong flaring of the
Sco-like sources. The presence of a full Z-track is also a feature of the Cyg-like sources.

The blackbody radius $R_{\rm BB}$ in the Cyg-like sources is generally larger than in the Sco-like sources
and has a maximum value on the Z-track excluding flaring (discussed later) at the soft apex
between 10 - 12 km as shown by the dashed lines. Although this is uncorrected it does suggest 
that emission is from the whole neutron star at the soft apex. On the NB and HB, 
$R_{\rm BB}$ decreases implying that the emission contracts to an equatorial belt on the neutron star.
It was proposed in our model for the Cyg-like group (Papers I, II and III)
that increasing radiation pressure of the neutron star on the NB and HB due to high $kT_{\rm BB}$
$\sim $2 keV disrupted the inner disk removing material above the orbital plane making
the disk thinner. This is expected to lead to accretion over a reduced equatorial part of the
neutron star. The details of the flow between inner disk and star are not well established,
however, there are two possibilities.
In the Inogamov \& Sunyaev model (1999) the accretion flow adjusts to the spin of the star
on the stellar surface producing X-ray emission from a strip of height that 
depends on $\dot M$. If accretion flow is diverted away from the neutron star, the emitting
area gets less. Alternatively, in a survey of LMXB with {\it ASCA}, the height of the emitting
strip was found to equal the height of the inner disk (Church \& Ba\l uci\'nska-Church 2001)
so that reduced disk height will mean reduced strip height. This equality could be explained
either by the Inogamov \& Sunyaev model, or possibly by some form of radial flow between disk 
and star.

In Sco\th X-1 and GX\th 349+2, a similar high $kT_{\rm BB}$ $>$ 2 keV is found at the
soft apex and the blackbody radius is reduced to a few km, which is consistent
with the Cyg-like source model. In GX\th 17+2, the blackbody radius is $\sim$ 8 km,
$kT_{\rm BB}$ being lower than in the other two sources.

The second major difference between the two groups is in flaring. Fig. 5 clearly
demonstrates the well-known fact that flaring in the Cyg-like sources is weak
but strong in the Sco-like sources. In Sco\th X-1 and GX\th 349+2, $R_{\rm BB}$ increases in flaring 
suggesting that the emitting area spreads from an equatorial belt across the neutron star. In the Cyg-like
sources there is similarly an increase of $R_{\rm BB}$ indicating that in this case the emission may expand 
outside the neutron star as in radius-expansion bursts.

The major spectral and implied physical changes in flaring
between the two groups are very apparent in the variation of luminosities of the individual blackbody
and Comptonized emission components shown next.

\subsection{Variation of the blackbody and Comptonized emission luminosities}

Fig. 6 shows the variation of the bolometric neutron star blackbody luminosity: $L_{\rm BB}$ and the 1 - 30 keV 
Comptonized emission of the ADC: $L_{\rm ADC}$ with the total luminosity $L_{\rm Tot}$ for
the Sco-like sources. These can provide valuable insights into the physical changes taking place. Firstly,
\begin{figure}[!ht]                                                      
\begin{center}
\includegraphics[width=84mm,height=84mm,angle=270]{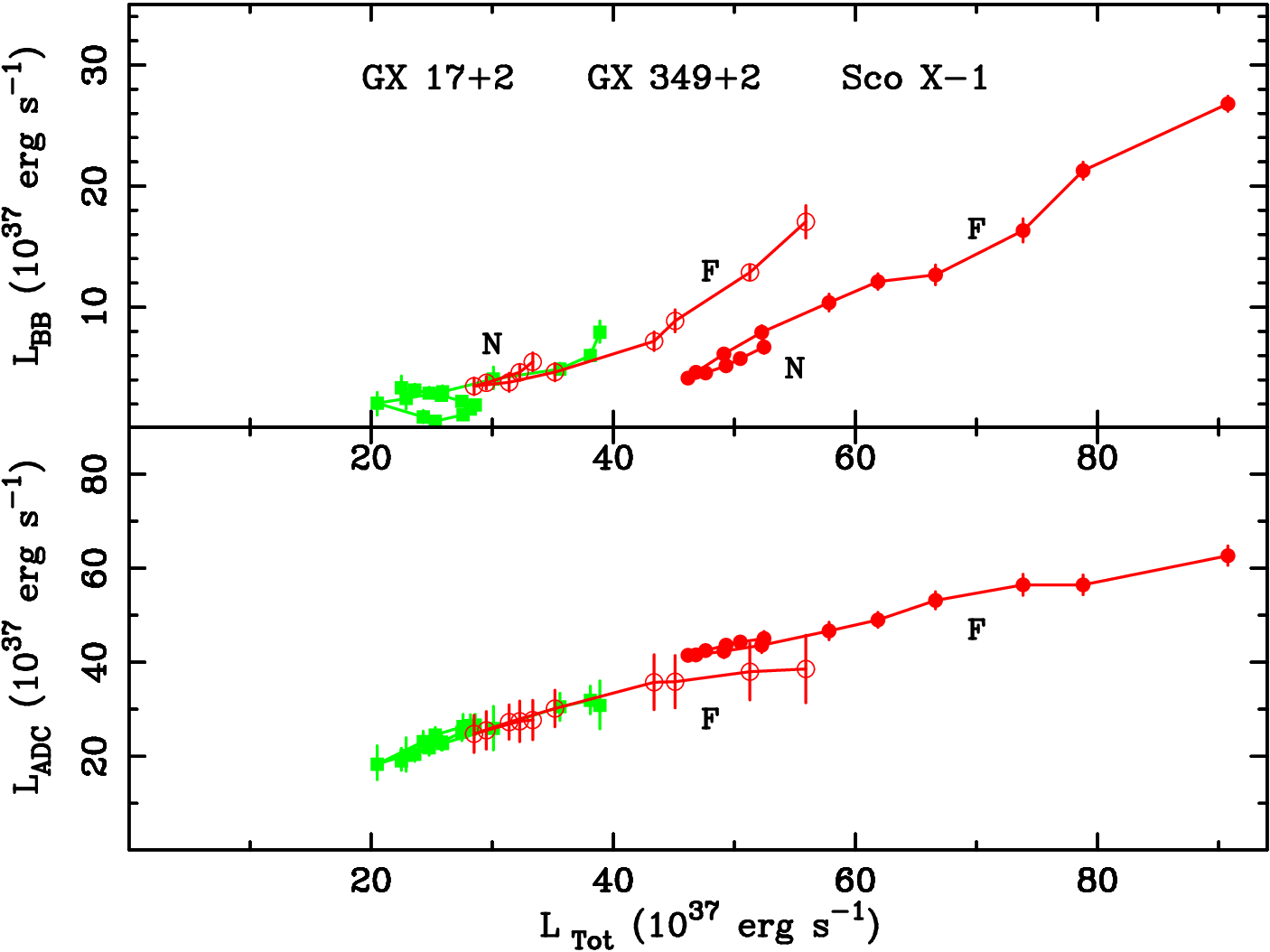}           
\vskip -76mm \hskip -10mm
\begin{minipage}[t]{40 mm}
\includegraphics[width=20mm,height=20mm,angle=270]{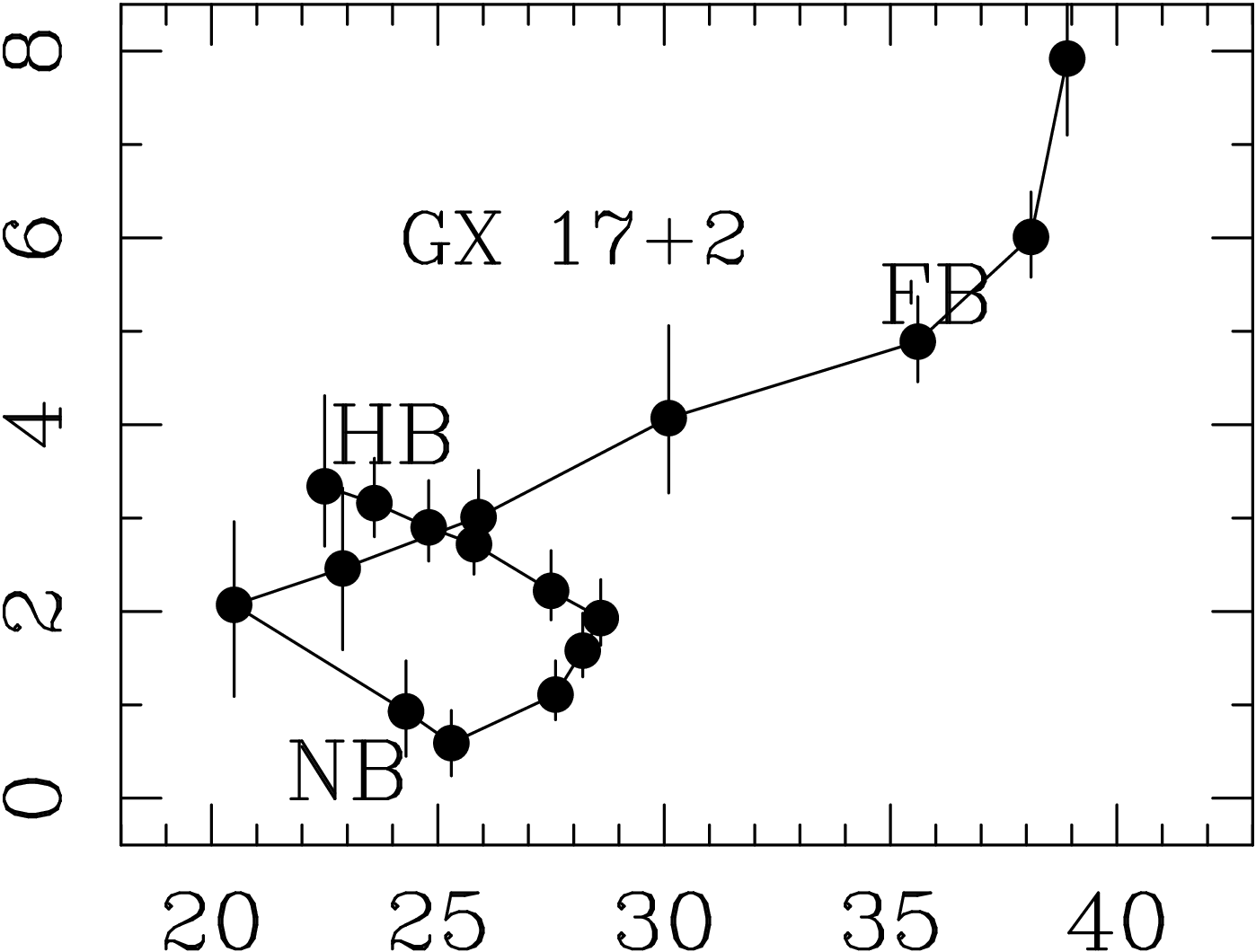}           
\end{minipage}
\vskip 70 mm
\caption{Variation of the blackbody luminosity (upper panel) and ADC luminosity 
(lower panel) around the Z-track in the Sco-like sources. To improve clarity, an expanded
view of GX\th 17+2 is given in the inset.}
\label{}
\end{center}
\end{figure}
\begin{figure}[!hb]                                                
\begin{center}
\includegraphics[width=84mm,height=84mm,angle=270]{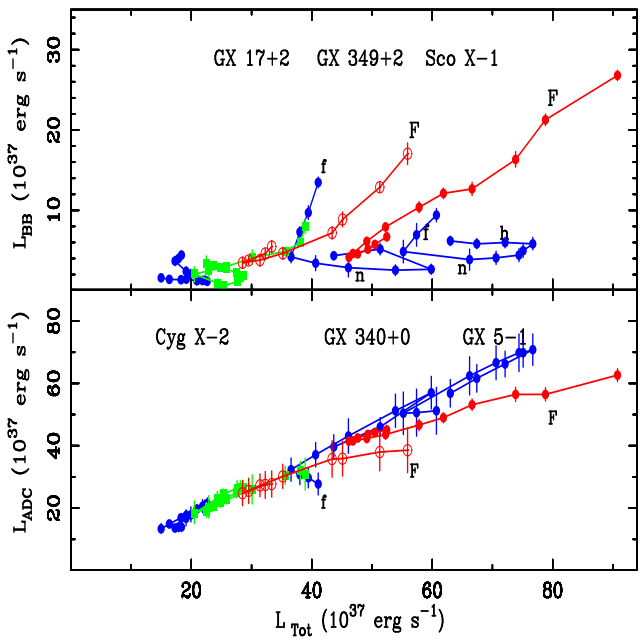}        
\caption{Comparison of the Sco\th X-1 like sources with the Cyg\th X-2 like sources.}
\label{}
\end{center}
\end{figure}
in all sources, $L_{\rm ADC}$ is dominant, the blackbody being only 10\% of the total luminosity 
at the soft apex, but increasing to $\sim$25\% at the peak of flaring in GX\th 349+2 and Sco\th X-1, 
and in these sources $L_{\rm BB}$ becomes itself super-Eddington at that point. 

In Fig. 7, the Sco-like sources are compared with the Cyg-like sources. In these, a large increase 
of the Comptonized emission luminosity is seen on the NB between the soft apex and the hard apex,
suggesting that the mass accretion rate $\dot M$ increases, based on the improbability of the ADC emission 
increasing strongly without $\dot M$ increasing. On the FB, $L_{\rm ADC}$ is constant within errors, 
while $L_{\rm BB}$ increases suggesting $\dot M$ is constant and the additional luminosity is generated 
on the neutron star, which we proposed was unstable nuclear burning (Paper I, II and III).

The Sco-like sources follow closely the same linear variation of $L_{\rm ADC}$ with $L_{\rm Tot}$ as 
expected given that $L_{\rm ADC}$ is the dominant emission in both sub-groups. 
However, there is a large increase of $L_{\rm ADC}$ in flaring, very different from the Cyg-like 
sources. There is also a large increase in $L_{\rm BB}$. This indicates that flaring in the Sco-like
sources may consist of unstable burning {\it combined} with $\dot M$ increase as proposed in the Discussion.

\subsection{Radiation pressure}

\begin{figure}[!h]                                                
\begin{center}
\includegraphics[width=84mm,height=84mm,angle=270]{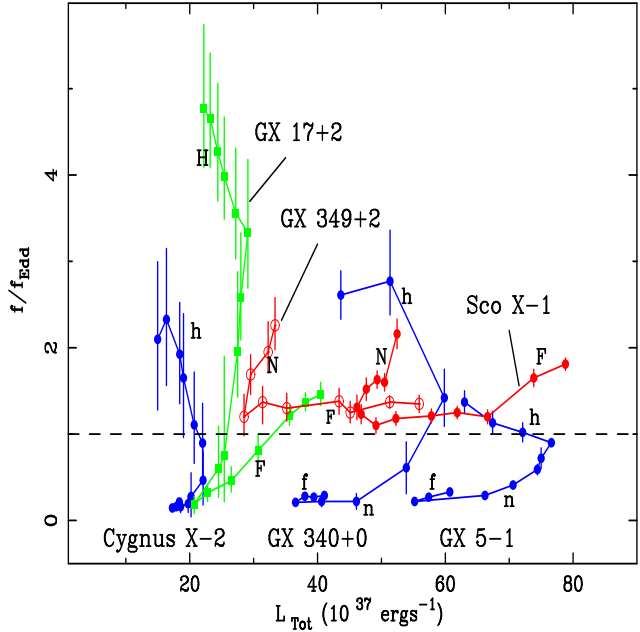}        
\caption{The emissive flux $f$ of unit area of the neutron star as a fraction of the Eddington
flux $f_{\rm Edd}$ for the Sco\th X-1 like sources compared with the Cyg\th X-2 like sources.}
\label{}
\end{center}
\end{figure}

In Fig. 8 we show the blackbody emissive flux of unit area of the neutron star $f$ obtained 
directly from spectral fitting and equal to $L_{\rm BB}/4\, \pi\, R_{\rm BB}^2$, so that
it is the flux of unit {\it emitting} area. In the figure, it is shown as a fraction of
the Eddington flux $f_{\rm Edd}$ = $L_{\rm Edd}/4\, \pi\, R_{\rm NS}^2$, where $R_{\rm NS}$
is the actual radius of the neutron star, with $f_{\rm Edd}$ $\sim 1.4\times 10^{25}$ erg cm$^{-2}$ s$^{-1}$ 
for an assumed 10 km radius. This ratio $f/f_{\rm Edd}$ expresses the strength of the radiation pressure 
in the neighbourhood of the neutron star showing whether the emission from the emitting part of the stellar 
surface is super-Eddington. It is more useful when the emission is not from the whole star than
$L/L_{\rm Edd}$ which shows whether the source as a whole appears super-Eddington
to a more distant observer and is thus an average. $f/f_{\rm Edd}$ can be substantially greater
than unity when $L/L_{\rm Edd}$ is not.

The behaviour of the Cyg-like sources (Paper I, II and III)
is characterised by low values of $f/f_{\rm Edd}$ at the soft apex with little change on the
weak flaring branches. However, on the NB the ratio rises to unity at the hard apex due to
the increase of $kT_{\rm BB}$ and increases further on the HB. As this is obtained directly
from spectral fitting it is clear that the radiation pressure close to the neutron star
becomes high as the ratio becomes several times super-Eddington. Disruption of the inner accretion 
disk should therefore be expected as proposed in our model of the Cyg-like sources.

In GX\th 349+2 and Sco\th X-1, it can be seen that $f/f_{\rm Edd}$ is
always greater than unity, as a consequence of the blackbody temperature never dropping
below 2 keV. Thus these main Sco-like sources have the major physical difference that
radiation pressure is always high which must have a strong effect on the inner regions of the binary.
GX\th 17+2 is more similar to the Cyg-like sources:
sub-Eddington at the soft apex becoming super-Eddington on the NB. As noted before, there is however, 
strong flaring as in the Sco-like sources. 

It is interesting to see that $kT_{\rm BB}$ (Fig. 5) has a value of $\geq $2 keV in Sco\th X-1 
and GX\th 349+2 at all times and in the Cyg-like sources at the hard apex, and in all these cases
$f$ $\sim$ $f_{\rm Edd}$ as a result. Thus the neutron star has about the Eddington temperature
$kT_{\rm Edd}$ $\sim$2 keV at these positions as might be expected. 

\subsection{Nuclear burning}

\begin{figure}[!ht]                                                
\begin{center}
\includegraphics[width=84mm,height=84mm,angle=270]{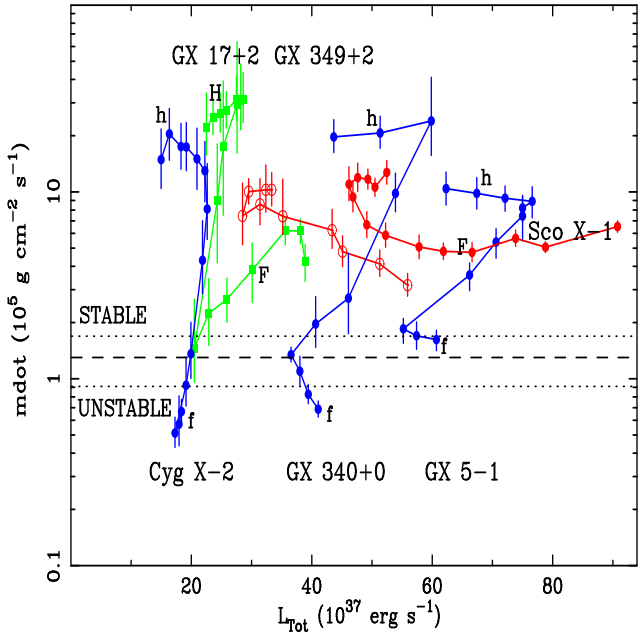}        
\caption{Values of the measured mass accretion rate per unit area to the neutron star compared
with the theoretical critical value $\dot m_{\rm ST}$ demarcating regimes of unstable and stable
burning (Bildsten 1998) shown as a dashed line. The dotted lines show the 30\% uncertainty attached
to the theoretical value.}
\label{}
\end{center}
\end{figure}

In Fig. 9 we compare conditions at the onset of flaring, i.e. at the soft apex, with the theory
of stable and unstable nuclear burning in the atmosphere of the neutron star. The critical parameter
in the theory is the mass accretion rate per unit area of the star $\dot m$ which for the wide range
of the total mass accretion rate $\dot M$ found in LMXB defines several r\'egimes of stable and unstable burning 
of H or He (Fujimoto et al. 1981; Fushiki \& Lamb 1987; Bildsten 1998; Schatz et al. 1999). 
$\dot m$ determines physical conditions in a column of the neutron star atmosphere above unit area
and so whether nuclear burning is stable or not. We evaluate $\dot m$ as the mass accretion rate 
per unit {\it emitting} area of the neutron star
given by $\dot m$ = $\dot M/ 4\,\pi \, R_{\rm BB}^2$ using $R_{\rm BB}$ from spectral fitting,
and $\dot M$ obtained from the total luminosity by standard accretion theory.

In the Cyg-like sources, it can be seen in Fig. 9 that the values of $\dot m$ at the soft apex
agree well with the critical theoretical $\dot m_{\rm ST}$ (Bildsten 1998) at the boundary of
unstable and stable He burning shown by the dashed line (with the 30\% theoretical uncertainty
shown as dotted lines). This provides evidence that flaring in these sources consists of unstable
nuclear burning (Paper I, II and III).

In Sco\th X-1 and GX\th 349+2, the values of $\dot m$ do not agree well with $\dot m_{\rm ST}$ and
are located some way into the stable region, several times higher than the critical value
(but not orders of magnitude away). It is clear that flaring in these sources has a different 
character which will be discussed below. GX\th 17+2 is again Cyg-like in this respect.

\subsection{ASM observations: the predominance of flaring in the Sco-like sources}

To help in a physical understanding of the differences between the Cyg and Sco-like sources, we
investigate quantitatively the differences in flaring behaviour using {\it RXTE} All Sky Monitor (ASM)
data. Lightcurves of all six sources were extracted over a $\sim$ 15-year period from 1996, Jan 6 to 2010, Nov 30
as shown in Fig. 10. Data were accumulated in 90 second dwells several times per day.
This immediately demonstrates the predominance of flaring in the Sco-like sources
(lower panel). In Cygnus\th X-2, the peaks visible are maxima in longterm variability at tens of days 
and not flares. The figure reveals extensive variability clearly unconnected with
Poisson noise. In Sco\th X-1, for a 90 second bin and 1000 count s$^{-1}$, Poisson variations
are 0.3\%; in the other sources with $\sim$50 count s$^{-1}$ the level
is 1.5\%. 
\begin{figure}[!h]                                                
\begin{center}
\includegraphics[width=44mm,height=84mm,angle=270]{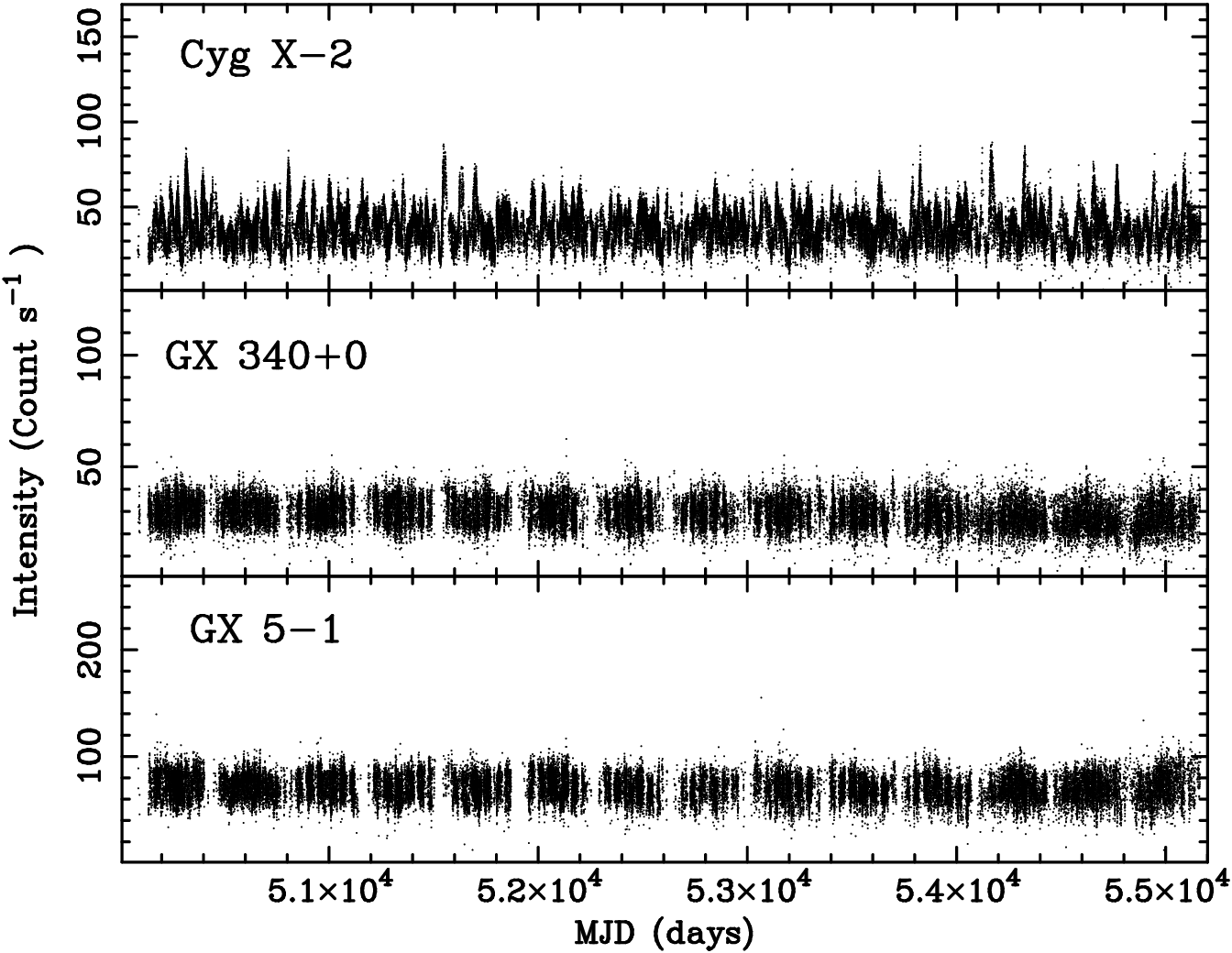}        
\includegraphics[width=44mm,height=84mm,angle=270]{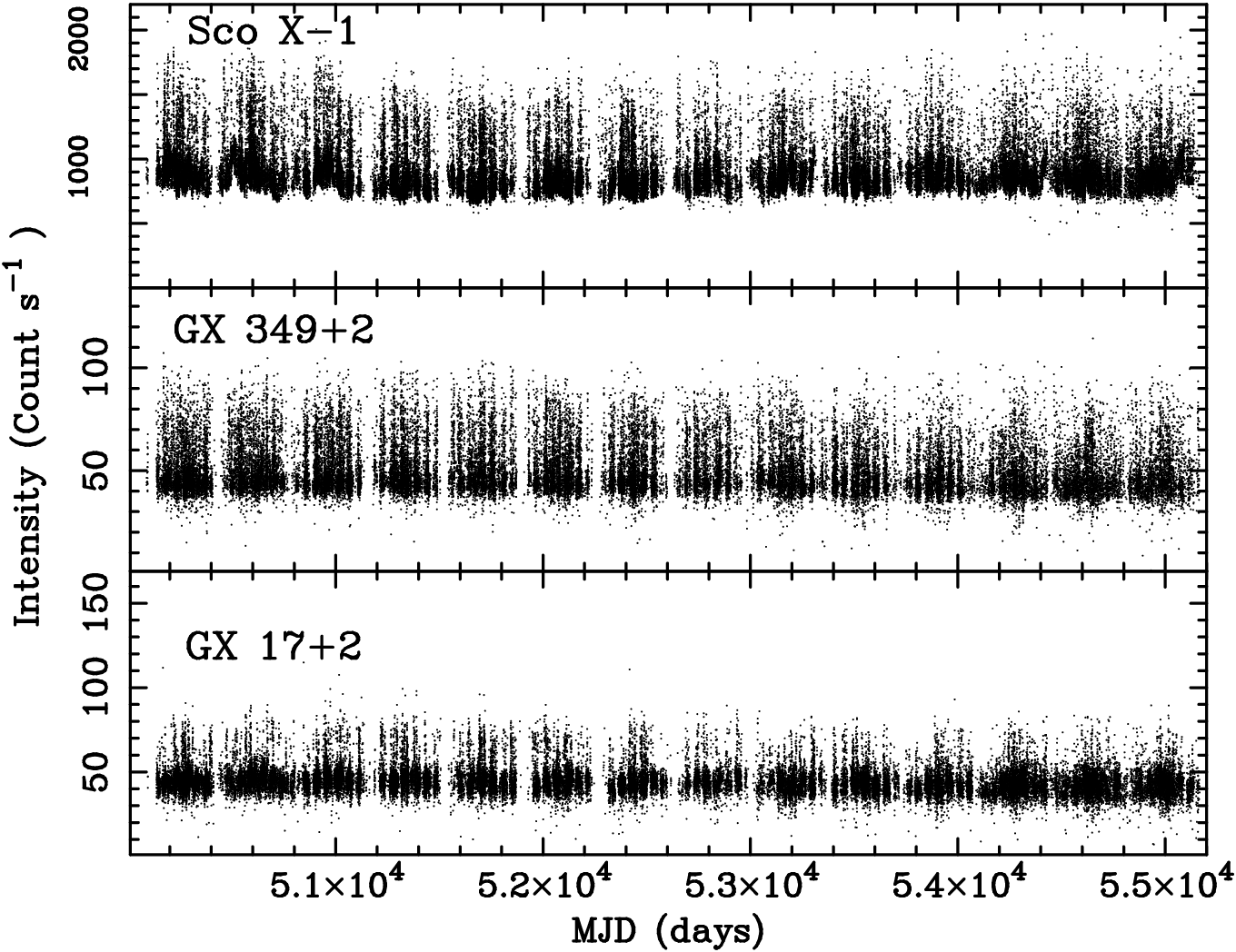}        
\caption{Comparison of flaring in the Sco-like and Cyg-like sources using 15 years of {\it RXTE} 
All SKy Monitor data. The peaks in the Cyg\th X-2 lightcurve are due to source variability,
not flaring (see text).}
\label{}
\end{center}
\end{figure}
\vskip  100mm

In Cyg\th X-2 we examined all pointed observations within the ASM lightcurve.
By extracting lightcurves and Z-tracks in hardness-intensity we identified position on the 
Z-track in these parts of the ASM data. Variations on a timescale of about 2 days were identified as 
Z-track movement. A  variation over about 40 days could also be seen possibly associated with secular 
motion of the Z-track (giving larger peaks in the 15-year lightcurve). Gaps in the ASM data reflect 
periods when the source had less coverage by the satellite. 
We were able to identify the hard apex in the Cyg-like sources as peaks in the two-day cycle and 
also in some cases the soft apex. However, the occurrence of flaring in pointed observations is rare
and it is difficult to find FB data without pointed observations as there is a limited 
increase in intensity. 

By contrast, in Sco\th X-1, the X-ray count rate at the peak of flares is much greater than 
at the hard apex, so the frequent peaks in the lightcurve (intensity $>$ 1100 count s$^{-1}$) 
can clearly be identified as flaring. The data were analysed using local software
to detect flares and identify the peak of each. Broader flares exhibit structure appearing to be 
a superposition of flares and it was assumed that peaks in the lightcurve less than 0.1 day apart 
represented structure and not the next flare. A lightcurve of detected flares (with their
appropriate count rates from the original lightcurve) was made. Careful examination of the flares 
detected in this way showed that the gaps between adjacent flares vary between $\sim$0.2 and $\sim$2.5 
days for the majority of flares. Only $\sim$40 flares had much larger spacings of 10 days in the
15 years of data. In GX\th 349+2 and GX\th 17+2, lower limit of 60 and 55 count s$^{-1}$
were applied in flare detection. In  GX\th 349+2, the majority of flares were spaced at less than 1.5 days
and few spacings ($\sim$5) of 10 days were seen. In GX\th 17+2, most flares were less than 4 days apart
but there were $\sim$85 flares with a spacing of 10 days. 

The total numbers of flares found in the 15 years of ASM data defined on the above criteria 
was 3013 in \hbox{Sco\th X-1,} 3523 in GX\th 349+2 and 1226 in GX\th 17+2. Dividing by the duration of the 
observations gives the mean flare rates and so the mean intervals between flares 
of 1.68 days, 1.44 days and 4.14 days in the three 
sources, respectively. There are about three times less flares in GX\th 17+2 as is obvious in the lightcurves 
supporting the idea that the source is transitional between Sco-like and Cyg-like. 

The results show clearly the strength of flaring in the Sco-like sources Sco\th X-1 and GX\th 349+2 
particularly. We will suggest (Sect. 4.2) that this causes the main physical difference of high
neutron star temperature in these sources. 

\subsection{Sco\th X-1 in a state of reduced flaring}

Finally we show results of analysis of the observation of Sco\th X-1 selected to have much reduced 
flaring. The lightcurve of this observation was shown in Fig. 2
and as noted earlier, the bulk of this observation consists of 110 ksec on the HB or NB 
without apparent flaring, and preceded by 120 ksec without strong flaring (not shown)
so that the data analysed are at least $\sim$150 ksec from any previous strong flare.
Fig. 3 shows the Z-track in hardness-intensity (lower central panel) which is remarkable in
having not only a normal branch but also a hard apex at about the end of the Z-track. 
The major flare at a time of $\sim$120 ksec seen in the Z-track occurred {\it after} the bulk 
of the observation on the HB and NB so did not affect spectra selected in those regions.

\begin{figure}[!h]                                                    
\begin{center}
\includegraphics[width=84mm,height=84mm,angle=270]{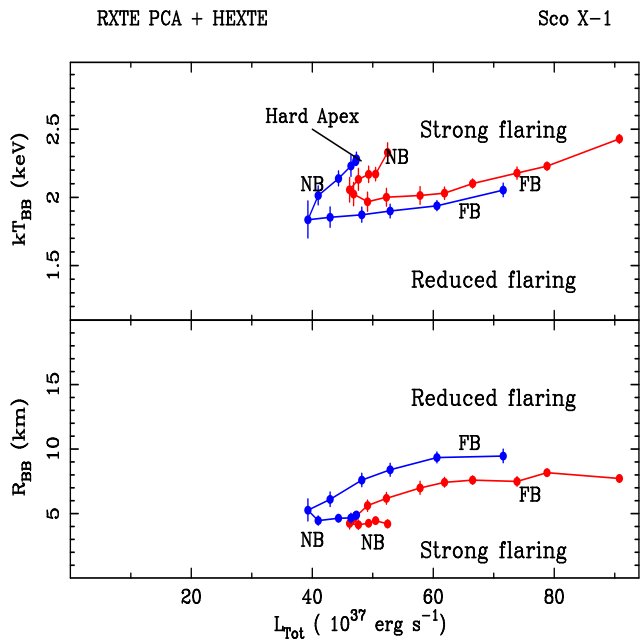}        
\caption{Comparison of blackbody temperature and radius for the two observations of Sco\th X-1,
one having strong flaring as normal, and the other during reduced flaring.
}
\label{}
\end{center}
\end{figure}

Spectra were selected along the Z-track and spectral analysis carried out as described previously.
Fig. 11 shows the most significant result: the blackbody temperature and blackbody radius 
compared with the observation of Sco\th X-1 with strong flaring. The blackbody temperature $kT_{\rm BB}$ 
increases on the normal branch as seen in all Z-track sources.

A reduction in $kT_{\rm BB}$ is seen in the observation with reduced flaring at all points on the flaring 
branch and the soft apex. Here, the value of $kT_{\rm BB}$ = 1.83$\pm$0.14 compared with 2.06$\pm$0.09 keV
in the flaring observation is significantly different at the 2$\sigma$ level. However, on
considering all of the FB data, the significance of the decrease in $kT_{\rm BB}$ becomes higher.
In many observations of Sco\th X-1 analysed by us, the value of $kT_{\rm BB}$ is more than 
2 keV so the value of 1.83 keV is quite unusual. The analysis provides evidence that the usual high value 
of $kT_{\rm BB}$ is associated with the frequent and strong flaring 
in this source, and we propose below that it is the strong flaring in the Sco-like sources that 
almost always maintains a high blackbody temperature giving rise to the observed spectral properties.

The blackbody radius $R_{\rm BB}$ increases in flaring in both observations increasing towards the 
neutron star radius at the peak of flaring, as noted previously by Barnard et al. (2001).    

\section{Discussion}

\subsection{Differences between the Sco\th X-1 like and the Cyg\th X-2 like sources}

\begin{figure}[!h]                                                    
\begin{center}
\includegraphics[width=84mm,height=84mm,angle=270]{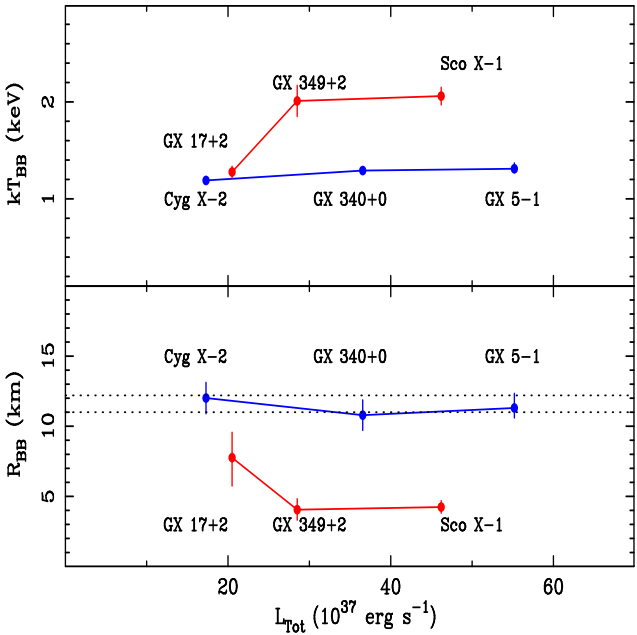}        
\caption{The major difference between the Sco\th X-1 like and Cyg\th X-2 like sources at the soft apex;
upper panel: the blackbody temperature $kT_{\rm BB}$; lower panel: the corresponding blackbody radii
$R_{\rm BB}$. The dotted lines correspond to the upper and lower limits of the average $R_{\rm BB}$
for the three Cyg-like sources of 11.6$\pm$0.6 km at 90\% confidence.}
\label{}
\end{center}
\end{figure}

On the normal branch, the Sco-like and Cyg-like sources are similar. In the latter, we suggested that the large 
increase of $L_{\rm ADC}$ was due to increasing mass accretion rate. $L_{\rm ADC}$ also increases in the Sco-like 
sources on the NB (Fig. 7) supporting the idea that $\dot M$ increases on the NB in all Z-track sources, unlike 
the often-held standard view that $\dot M$ increases monotonically in the opposite direction 
HB$\rightarrow$NB$\rightarrow$FB.

The neutron star temperature in the Cyg-like sources is lowest at the soft apex and the blackbody radius maximum 
(for the non-flaring source) suggesting that the whole star is emitting. The average value for the 3 sources is 
11.6$\pm$0.6 km, but should not be taken literally as the neutron star radius. Firstly, a relativistic correction 
(1 + z) of 34\% would reduce the value to 8.7 km. Also applying a colour correction (Sect. 3.4) would
however, increase the final value to 18.7$\pm$0.97 km which is disturbingly high. In addition there may be possible 
systematic effects such as blocking of part of the neutron star 
emission by the disk or corona, although this is not generally seen in LMXB except in dipping. Cygnus\th X-2 is 
the only source with dipping and high inclination; however, $R_{\rm BB}$ at the soft apex in this source does not appear
different from the other Cyg-like sources.

It was proposed that increasing radiation pressure on the NB disrupted the inner disk leading to accretion on an equatorial belt,
and possible mechanisms for this suggested (Paper I). 

In Sco\th X-1 and GX\th 349+2, $kT_{\rm BB}$ is substantially higher, this constituting the first major difference, 
and $R_{\rm BB}$ is reduced. To emphasize the differences in Fig. 12 we show $kT_{\rm BB}$ and $R_{\rm BB}$ for all 
6 sources at the soft apex. The reduction of $R_{\rm BB}$ in the Sco-like sources at the soft apex with higher $kT_{\rm BB}$ 
is consistent with the reduction in the Cyg-like sources with increasing $kT_{\rm BB}$ on the NB.

Closely related is the radiation pressure of the neutron star emission expressed via $f/f_{\rm Edd}$ using 
the measured flux (Fig. 8). In Sco\th X-1 and GX\th 349+2 the values are always super-Eddington whereas 
the Cyg-like sources have sub-Eddington values at the soft apex becoming high at the hard apex where jets 
are observed. 
Thus in \hbox{Sco\th X-1} and GX\th 349+2 in which the radiation pressure is always high we expect the 
sources to be permanently subject to disruption of the inner disk and jet launching may take place even from
the soft apex although $f/f_{\rm Edd}$ is only unity. Whether this occurs requires simultaneous radio and X-ray 
observations during Z-track movement. We might also possibly expect jets during flaring on the basis that 
$f/f_{\rm Edd}$ remains high. 

GX\th 17+2 appears transitional not having a high $kT_{\rm BB}$ at the soft apex nor a very reduced  $R_{\rm BB}$. 
Flaring is however, similar to that in Sco\th X-1 and GX\th 349+2 with a definite increase of Comptonized emission.

The second major difference is in flaring. The Cyg-like sources have constant $L_{\rm ADC}$ arguing for a 
constant $\dot M$ but an increase of blackbody luminosity. In the Sco-like sources, $L_{\rm ADC}$ increases
and the FB appears more similar to the NB (Fig. 7). In the Cyg-like sources, we proposed
an additional energy source of unstable nuclear burning on the neutron star giving increased $L_{\rm BB}$ at
constant mass accretion rate. This was supported by good agreement of the mass accretion rate per unit area 
$\dot m$ at the soft apex with the theoretical $\dot m_{\rm ST}$ at the border of unstable/stable
He burning in bright LMXB in a mixed H/He environment (e.g. Bildsten et al. 1998). 
For a source 
descending the NB, at the soft apex conditions in the neutron star atmosphere allow unstable burning and 
a flare begins. The Sco-like sources are clearly different with increases in both $L_{\rm ADC}$ and 
$L_{\rm BB}$. This suggests an increase of $\dot M$ on both NB and FB, while some other factor must be different 
in flaring and it is natural to think that this may be unstable burning. The increase of blackbody luminosity
in flaring in the Sco-like sources is similar to that in GX\th 340+0 (Fig. 7) supporting this. In all Z-track sources
there is an increase of blackbody radius in flaring (Fig. 5) suggestive of spreading of the burning
region as might be expected across the neutron star in the Sco-like sources and possibly in an expanding
atmosphere in the Cyg-like sources. 

In GX\th 17+2, $\dot m$ at the soft apex is close to $\dot m_{\rm ST}$ (Fig. 9) so that unstable 
burning is expected in flaring. However, in Sco\th X-1 and GX\th 349+2 $\dot m$ $>$ $\dot m_{\rm ST}$ 
implying that burning will be stable. A possible explanation is that with high radiation pressure at all times
in these sources, the mass flow to the neutron star is modified by 
disruption of the inner disk, with mass permanently diverted out of the disk. $\dot m$ is obtained 
from $\dot M$ assuming the standard relation with total luminosity, so that
if mass is diverted away from the neutron star we over-estimate 
$\dot M$ but $\sim$80\% of the flow would have to be diverted to bring $\dot m$ into the 
unstable region.

\subsection{A model for the Sco\th X-1 like sources}

The results suggest a model for the Sco-like sources as follows. Ascending the NB the sources are similar to
the Cyg-like sources with $\dot M$ increasing. Flaring is very different as $\dot M$ also increases suggesting 
that flaring is unstable nuclear burning {\it combined} with increasing $\dot M$. The additional burning process
causes the FB in hardness-intensity to deviate from the NB (becoming less hard). Variations of the mass
accretion rate sometimes cause the source to move on the NB, but if unstable burning is triggered
then the movement is on the FB. Unstable burning might be triggered if part of the mass flow has been 
diverted into the jet and away from the neutron star so that $\dot m$ $<$ $\dot m_{\rm ST}$. The radiation
pressure is higher on the NB (Fig. 8) in the Sco-like sources so a source descending the NB
may fulfil the unstable burning condition. 
Another possibility is that if only part of the neutron star is emitting at the soft apex, on the edge of the 
emitting region $\dot m$ may be reduced and unstable burning possible.
Galloway (2008) made a similar suggestion that accretion over a limited part of the star might
at lower mass accretion rates produce X-ray bursts confined to part of the star, which might spread
across the star (Bildsten 1995). The Cyg-like sources at the soft apex accrete over the whole star
apparently so unstable burning only occurs when $\dot m$ falls below the critical value. 
The Sco-like sources would be expected to move onto the NB from the soft apex after a 
period of flaring as further accumulation of matter would be needed for the next flare.


Secondly, the high temperature of the neutron star in Sco\th X-1 and GX\th 349+2 at all times 
$>$ 2 keV results from the continual strong flaring releasing energy on the neutron star
in contrast with the Cyg-like sources where flaring is rare and weak. Between flares, the neutron star could 
cool in principle by the emission of X-rays; however, it appears observationally that cooling is not effective 
on the timescale between flares as discussed below.

Thirdly, the model predicts that jets are formed when the radiation pressure is high. In the Cyg-like sources,
this is in the region of the hard apex but not at the soft apex. However, in Sco\th X-1 the radiation pressure
is high even at the soft apex and in flaring. The model thus predicts that radio emission in Cyg\th X-2 will sometimes be zero,
i.e. quenched in flaring but emission in Sco\th X-1 may not become zero at the soft 
apex. In Cyg\th X-2 radio fluxes up to 5 mJy were found on the horizontal and upper normal branch at 1.49, 4.9 
and 8.4 GHz but the source was not detected on the lower NB and FB with flux  $<$ 1 mJy, i.e. showing a clear 
correlation with Z-track position (Hjellming et al. 1990a). Similarly, Ba\l uci\'nska-Church et al. (2011) found
a 5$\sigma$ upper limit of $<$ 150 $\mu$Jy at 5 GHz in e-EVN observations in 2009, May. Sco X-1 on the other hand 
has both radio-loud and radio-weak states and weak radio emission has been detected when on the FB 
(Hjellming et al. 1990b). This tends to support the prediction but clearly further testing is needed.

\subsection{Cooling rate}

We propose that unstable nuclear burning deposits energy in the neutron star such that, between 
flares, i.e. at the soft apex, the blackbody temperature measured for the atmosphere does not
decrease but remains at 2 keV as in a flare. This may be compared with X-ray bursting in which
the emission cools rapidly from $\sim$ 2.2 keV to 1 keV between bursts. It thus appears that radiative
losses are not effective in cooling the neutron star between flares. However, we found some evidence
for a 10\% cooling after 3 days without flares. This raises the interesting question of 
the rate of cooling of the neutron star. 

In the case of X-ray burst 
sources, a small fraction of the nuclear energy released will flow into the interior at each burst, heating 
it. The time-averaged inward flow will be small: $\sim$ 4\% (Hanawa \& Fujimoto 1984), because 
most of the energy released is radiated. In equilibrium, there will be an 
outward energy flow balancing this (Lamb \& Lamb 1978; Ayasli \& Joss 1982; Hanawa \& Fujimoto 1984).
Compressional heating and neutrino cooling will also be involved.
The averaged inward flow
is small, $\approx 10^{35}$ erg s$^{-1}$, and the thermal capacity of the neutron star dominated by 
partially degenerate electrons is large, assuming the neutrons are 
super-fluid (Tsuruta et al. 1972) $\approx$ $3\times 10^{44}$ erg (Lamb \& Lamb 1978). Thus the 
heating timescale can be $\sim$100 years. 

Unstable burning in flaring will also heat the star but whereas a burst lasts $\sim$100 seconds 
and occupies only 1\% of the interval between bursts, flaring is more prolonged at 
$\sim$5000 seconds and the ratio of flare duration to flare spacing is higher. Thus the inward heat
flow will be higher and so the heating timescale reduced, and in equilibrium, the outward 
flow of energy from the interior would also be increased. 

To compare with our results, we consider what happens if the flaring 
is suddenly turned off after a long period
of continuous flaring. If we assume an outward flow of heat much higher than in
the burst source case, say $10^{37}$ erg s$^{-1}$, for the whole neutron star to cool
(i.e. the partially degenerate electrons) by 10\% in temperature would require 
$\approx$0.1 year, about 10 times longer than the probable 3 days timescale. 
This may suggest that the appropriate cooling rate is not the equilibrium rate of
outward heat flow, but a higher rate, for example, electrons in the crust may cool
more rapidly than that.

\subsection{The energetics of flaring}

It may be thought that since nuclear burning, e.g. of hydrogen, is 30 times less efficient than
accretion power, flaring cannot be nuclear in nature, and in the standard view of Z-track sources
it is increasing $\dot M$ that powers flares in both Cyg and Sco-like sources. However, in the Cyg-like
sources an increase of neutron star luminosity in flaring of $10\times 10^{37}$ erg s$^{-1}$ for 2000 seconds
say, releases a flare energy $E_{\rm flare}$ of $\sim 1\times 10^{41}$ erg. Instantaneous nuclear burning
of the accretion flow cannot supply this, but matter accumulating over a period of $\approx$ 3 hours 
provides an energy $\epsilon\,M\,c^2$ $\geq$ $E_{\rm flare}$. The flares in the Sco-like sources are stronger
but only $\sim$50\% or less of the energy appears to be nuclear burning. $L_{\rm BB}$ increases by $\sim 20\times 10^{37}$ 
erg s$^{-1}$  giving $E_{\rm flare}$ $\sim 1\times 10^{41}$ erg which again can be provided by nuclear burning
of accumulated matter.

This is completely analogous to nuclear burning in X-ray bursts in which it was realized at an early stage
that although the energy release per nucleon is much less than the accretion release, if matter is accumulated
and consumed suddenly, the nuclear burning temporarily dominated the energy release (Lamb \& Lamb 1078).

\subsection{Does luminosity determine Cyg-like or Sco-like nature ?}

The transient source XTE\th J1702-462 having an outburst in 2006-7 (Remillard et al. 2006) rose to high
luminosity at the start of the outburst and exhibited Cyg-like properties, but as the brightness decreased
became more similar to Sco-like sources and eventually displayed Atoll source properties (Homan et al. 2007; 
Lin et al. 2009). Homan et al. (2010) thus proposed that luminosity, i.e. mass accretion rate determined 
Cyg-like or Sco-like behaviour, higher $\dot M$ making the source Cyg-like. As pointed out by Homan et al., 
this also means that differences in inclination or magnetic field strength cannot determine the nature 
as previously suggested. The suggestion that luminosity determines Cyg or Sco-like behaviour is not supported 
by our results as we see no systematic differences of luminosity between the two types as shown in the figures, 
the Cyg-like sources spanning a factor of three in luminosity and the Sco-like sources almost as much. (It has 
been previously widely accepted, of course, that Cyg\th X-2 was less luminous than Sco\th X-1, for example).

It could be that distance errors lead to the apparent lack of dependence of source type on luminosity. For luminosity 
to determine type, {\it all} the Cyg-like sources would have to be more luminous than {\it all} the Sco-lke sources.
In particular, Cygnus\th X-2 must be more luminous than Sco\th X-1 for which the distance
is well-determined at 2.8$\pm $0.3 kpc. The distance of Cyg\th X-2 assumed here to be 9.0 kpc (Table 6) would have 
to be  $>$ 14.7 kpc for the source to be as luminous as Sco\th X-1 ($46\times 10^{37}$ erg s$^{-1}$ at the soft apex).

\tabcolsep 5.0 mm
\begin{table}[!h]
\begin{center}
\caption{Source distances used in analysis}
\begin{minipage}{84mm}
\begin{tabular}{lr}
\hline \hline
source               &distance\\
                     & (kpc)\\
\noalign{\smallskip\hrule\smallskip}
GX\th 349+2          &9.2\\
Scorpius\th X-1           &2.8\\
GX\th 17+2           &7.5\\
Cygnus\th X-2        &9.0\\
GX\th 340+0          &11.0\\
GX\th 5-1            &9.0\\
\noalign{\smallskip}\hline
\end{tabular}\\
\end{minipage}
\end{center}
\end{table}

Distances to Cyg\th X-2 were published by Cowley et al. (1979) of 5.7 - 8.7 kpc for assumed donor masses of 
0.5 - 1.0 M$_{\sun}$. Orosz \& Kuulkers derived a donor mass of 0.6$\pm$0.13 M$_{\sun}$ and a distance of 7.2$\pm$1.1
kpc. From peak burst fluxes, Smale (1998) found 11.6$\pm$0.13 kpc and Galloway (2008) 11$\pm$2 kpc for an H fraction 
X = 0.7 and 14$\pm$3 for X = 0; however, not all bursts exhibited the characteristics of photospheric radius expansion, 
moreover bursting in the Z-track sources Cyg\th X-2 and GX\th 17+2 is not fully
\begin{figure}[!hb]                                                    
\begin{center}
\includegraphics[width=34mm,height=84mm,angle=270]{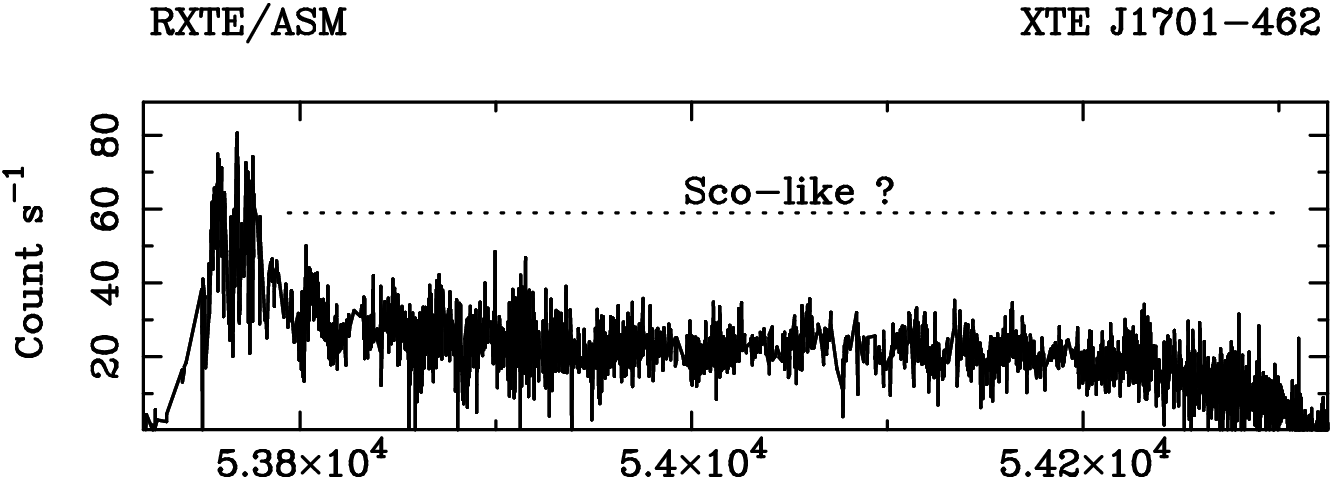}        
\includegraphics[width=32mm,height=84mm,angle=270]{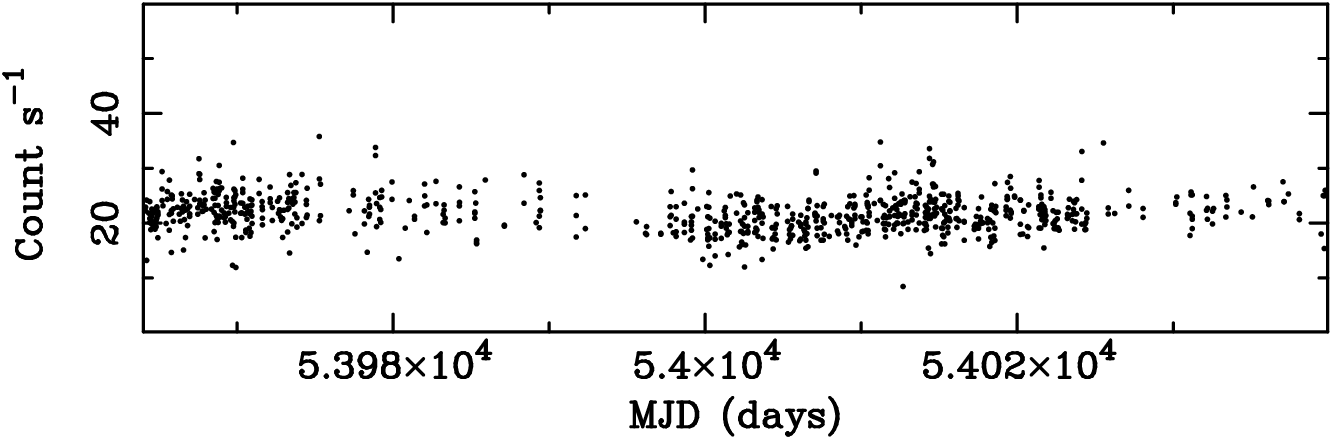}        
\caption{ASM lightcurve of the transient XTE\th J1702-462: \break upper panel: the complete outburst; lower panel:
expanded view showing the limited extent and occurrence of flaring}
\label{}
\end{center}
\end{figure}
understood. Jonker \& Nelemans (2004) give
11.4 - 15.3 kpc for hydrogen rich to helium rich material. Thus the distance is not well-constrained and Cyg\th X-2 could 
be {\it as luminous} as Sco\th X-1. However, to produce properties very different from Sco\th X-1 it would have to be 
appreciably {\it more luminous} needing a distance more than 20 kpc and this is unlikely.

For GX\th 340+0, distances have been found of 10.2 kpc (Penninx 1989), 11.8 kpc (van Paradijs \& White 1995) and 11.0 kpc 
(Grimm et al. 2002). $L$ is $36.6\times 10^{37}$ erg s$^{-1}$ at the soft apex for the assumed 
11 kpc. This would need to be 12.3 kpc to be as luminous as Sco\th X-1 which is clearly possible, but $\sim$17 kpc to 
be twice as luminous, say, which appears unlikely.

The identification of Sco-like behaviour in the transient was based on the shape of the Z-track. In the notation
of Lin et al. (2009) the transient was Cyg-like in section I of the lightcurve and Sco-like in sections II and III.
The appearance of hardness-intensity in II was similar to GX\th 17+2 with a long NB; however, we suggest that
this source is not fully Sco-like but transitional between the Cyg and Sco types. In III, the NB becomes small and the FB large and the source does appear Sco-like.
However, the intensity increase in flaring was \hbox{$\leq$ 40\%} in the total PCA band while all Sco-like sources
have 100\% increases or more (Figs. 2 and 10), and Cyg\th X-2 sometimes displays flares of up to 50\%, so the
flaring is not conclusive. In several respects, Homan et al. (2007) also did not find XTE\th J1702-462 unambiguously 
Sco-like.

Towards resolving whether the transient source was truly Sco-like, in Fig. 13 the ASM lightcurve of the transient source 
is shown for comparison with the lightcurves of Sco\th X-1 and GX\th 349+2 (Fig. 10). The upper panel shows the complete
brightening and decay of the transient; the dotted line shows the part which was thought to be similar to the Sco-like
sources; a known modulation at 25 days (Homan et al. 2007) can be seen. In the lower panel, a restricted part of the 
lightcurve is shown so that the extent of flaring can be seen without this modulation. It is evident that the flaring is 
not continuous or strong as in Sco\th X-1 but sparse and weak suggesting that the source was not Sco-like.

Rather than making identification on the appearance of the Z-track, based on the present work we can offer a physical 
definition of Sco-like nature: that the neutron star blackbody temperature $kT_{\rm BB}$ is at all times $>$ 2 keV, 
and that there is frequent strong flaring. Thus GX\th 17+2 is not fully Sco-like but transitional between Cyg and Sco-like. 
We suggest that XTE\th J1702-462 may not have been Sco-like as it decayed but remained Cyg-like with stronger flaring;
this is sometimes but not often seen in Cyg\th X-2. On the basis 
that the source would move closer to the condition for unstable burning ($\dot m$ $<$ $\dot m_{\rm ST}$) as luminosity 
decreased, this would be expected. Whether the transient source satisfied the above definition for being Sco-like, 
or not, could be checked by spectral analysis. If the source did not become Sco-like, then luminosity 
may not determine whether sources are Cyg-like or Sco-like.

\subsection{A general Z-track model}

Application of the Extended ADC model having the form discussed in Sect. 1.4 has provided
a physical model for the Z-track sources.
In summary, in all types of Z-track source, ascending the normal branch is due to an increase of $\dot M$
for which spectral fitting presents convincing evidence, as a definite increase of both total
X-ray luminosity $L_{\rm Tot}$ and ADC luminosity $L_{\rm ADC}$. The resultant
high neutron star temperature gives a high radiation pressure launching the jets. In Cyg\th X-2
this predicts an intermittent jet when the source is on that part of the Z. In Sco\th X-1 
the jet may be present even at the soft apex due to the permanently high $kT_{\rm BB}$.

Flaring in the Cyg-like sources consists of unstable nuclear burning. In the Sco-like sources, 
$\dot M$ increases as the source moves away from the soft apex on the NB or FB.
Sometimes conditions favour unstable burning so making the FB distinct from the NB. Strong flaring
in the Sco-like sources causes the observed spectral differences, i.e. high neutron star temperature 
and associated radiation pressure. The permanently high $kT_{\rm BB}$ and high radiation pressure 
may suggest a permanent bending of the accretion flow out of the plane into the jet. The 
residual flow to the neutron star may then satisfy the condition for unstable burning ($\dot m$ $<$ 
$\dot m_{\rm ST}$) so that a period of matter accumulation is followed by flaring, this then repeating. 
In the Cyg-like sources there is no permanent bending of the accretion flow, so unstable burning only 
takes place when $\dot M$ is low. The relevant question then becomes what is the fundamental cause of 
the strong flaring in the Sco-like sources.

As discussed in Sect. 4.5, the present results do not confirm that Cyg or Sco-like nature is
determined by luminosity as suggested by Lin et al. (2009). In this case, some other factor is responsible.
Flaring in the Cyg-like sources will only take place when a decrease of $\dot M$ causes $\dot m$ 
to fall below the critical value, and this makes flaring weak and infrequent. The fully Sco-like sources
(excepting GX\th 17+2) are clearly different with longterm diversion of accretion flow into jets 
implied. It is possible that this could
be connected with the long orbital periods of 18.9 and 21.85 hr in Sco\th X-1 and GX\th 349+2 as this 
suggests evolved companions and higher mass accretion rates. This could heat the neutron star
causing jet launching and conditions for unstable burning may become more often satisfied.

\thanks{
We thank the referee for his helpful comments.
This work was supported in part by the Polish KBN grants 3946/B/H0/2008/34 and 5843/B/H03/2011/40.}

\end{document}